\RequirePackage{lineno}
\documentclass[superscriptaddress,aps,prl,preprint,amsmath,amssymb,floatfix]{revtex4-2}
\usepackage{calrsfs}
\usepackage{graphicx,morefloats}
\usepackage{subfigure} 
\usepackage{color,soul}
\usepackage{comment}
\usepackage{amsmath}
\usepackage{amssymb}
\usepackage{times}
\usepackage{hyperref}
\usepackage{bbm}
\usepackage{soul}

\newcommand{\MT}[1]{{\color{black} #1}}

\begin{document}

\hyphenation{va-ni-sh-ing li-quid}

\title{Boosting the surface conduction in a topological insulator}

\author{M.~Taupin}
 \email{taupin@ifp.tuwien.ac.at}
 \affiliation{These authors have contributed equally.}
 \affiliation{Institute of Solid State Physics, TU Wien, Wiedner Hauptstr. 8-10, 1040 Vienna, Austria}
\author{G.~Eguchi}
 \affiliation{These authors have contributed equally.}
 \affiliation{Institute of Solid State Physics, TU Wien, Wiedner Hauptstr. 8-10, 1040 Vienna, Austria}
\author{M.~Lu\v{z}nik}
 \affiliation{Institute of Solid State Physics, TU Wien, Wiedner Hauptstr. 8-10, 1040 Vienna, Austria}
\author{A.~Steiger-Thirsfeld}
 \affiliation{USTEM, TU Wien, Wiedner Hauptstr. 8-10, 1040 Vienna, Austria}
\author{Y.~Ishida} 
\affiliation{ISSP, The University of Tokyo, Kashiwa, Chiba 277-8581, Japan}
\author{K.~Kuroda} 
\affiliation{ISSP, The University of Tokyo, Kashiwa, Chiba 277-8581, Japan}
 \affiliation{Graduate School of Advanced Science and Engineering, Hiroshima University, 1-3-1 Kagamiyama, Higashi-Hiroshima 739-8526, Japan}
 \affiliation{International Institute for Sustainability with Knotted Chiral Meta Matter (WPI-SKCM2), 1-3-2 Kagamiyama, Higashi-Hiroshima 739-8511, Japan}
\author{S.~Shin} 
\affiliation{ISSP, The University of Tokyo, Kashiwa, Chiba 277-8581, Japan}
\author{A.~Kimura}
 \affiliation{Graduate School of Advanced Science and Engineering, Hiroshima University, 1-3-1 Kagamiyama, Higashi-Hiroshima 739-8526, Japan}
 \affiliation{International Institute for Sustainability with Knotted Chiral Meta Matter (WPI-SKCM2), 1-3-2 Kagamiyama, Higashi-Hiroshima 739-8511, Japan}
\author{S.~Paschen}
 \email{paschen@ifp.tuwien.ac.at}
 \affiliation{Institute of Solid State Physics, TU Wien, Wiedner Hauptstr. 8-10, 1040 Vienna, Austria}

\date{\today}

\begin{abstract}
The protected surface conduction of topological insulators is in high demand for the next generation of electronic devices. What is needed to move forward are robust settings where topological surface currents can be controlled by simple means, ideally by the application of external stimuli. Surprisingly, this direction is only little explored. In this work we demonstrate that we can boost the surface conduction of a topological insulator by both light and electric field. This happens in a fully controlled way, and the additional Dirac carriers exhibit ultra-long lifetimes. We provide a comprehensive understanding, namely that carriers are injected from the bulk to the surface states across an intrinsic Schottky barrier. We expect this mechanism to be at play in a broad range of materials and experimental settings.
\end{abstract}

\maketitle

\noindent \section*{Introduction}
\noindent The past decade has seen a wealth of studies on topological insulators \cite{Hasan2010,Qi2011,Wehling2014}, motivated at least in part by the potential of these materials for the next generation of electronic devices. Despite massive efforts, unambiguously identifying the topological response in a given system remains challenging due to the difficulty to disentangle it from trivial effects. A direction that has been only little explored, despite its potential, is persistent stimulated conductivity: Photoconductivity studies \cite{Yeats2015,Zheng2015,Pirralho2019,Kang2019,Hou2020,Xie2021} indicate a strong response upon light excitation, but the actual role of the topological states is disputed. Seemingly disconnected studies---without the application of external stimuli---have reported similar persistent features with non-exponential relaxation \cite{Tian2017,Li2017,Uchiyama2023}, suggesting that they might have a common origin. Our work provides compelling evidence for this to be the case and identifies the underlying mechanism. We demonstrate that the electrical resistivity is strongly affected by external excitation (such as thermal radiation, light illumination, and current driving): Excited electrons migrate to the surface states and remain there ``permanently'' due to an intrinsic Schottky barrier and space-charge separation between the surface and bulk carriers. This leads to a significant enhancement of the surface conduction, which can be adjusted by the amplitude of the external excitation. All our transport data, both on bulk and microstructed samples, as well as our spectroscopic results, are consistently understood within this simple model. This also holds true for previously reported transport \cite{Yeats2015,Zheng2015,Tian2017,Li2017,Uchiyama2023,Pirralho2019,Kang2019,Hou2020,Xie2021} and spectroscopic \cite{Ishida2015,Neupane2015,Sumida2017,Yoshikawa2018,Papalazarou2018,Yoshikawa2019,Ciocys2020} observations and should be valid for any topological insulator. The control techniques we have demonstrated represent a proof-of-concept demonstration of controlled surface conduction in topological insulators. It may be exploited both in further fundamental studies and for applications.

\noindent \section*{Experimental details}
Our electrical transport study was performed on the bulk topological insulator Tl$_{1-x}$Bi$_{1+x}$Se$_{2-\delta}$ \cite{Eremeev2010, Lin2010, Yan2010, Sato2010, Kuroda2010}. 
 \MT{The single crystals were grown as described in Ref.\citenum{Kuroda2015}.

The resistivity was measured with spot-welded or silver-paint contacts using 10 or 25~$\mu$m diameter gold wires. Similar results were observed in different samples and with different contact configurations, highlighting the robustness of the results. The samples were obtained by cleaving a larger piece and were typically 0.5-1~mm in length and width, and 50-200~$\mu$m in thickness.
 The measurements were performed down to 2~K, in part in a Physical Property Measurement System (PPMS) from Quantum Design Inc.~using either the standard resistivity option or the ac-resistance bridge option for the high-current measurements, and in part in an Oxford flow cryostat with a home-made setup. In the latter, a lock-in SR830 and, for the current-voltage characteristics, a Keithley source meter 3624B were used. 

In the PPMS, the sample faces either the bottom of the last radiation buffer or the cap at the top of the sample chamber, depending on whether or not the radiation buffers provided by Quantum Design Inc. were installed. Accordingly, black-body radiation comes either from the temperature of the last buffer (thus defined as ``incomplete thermal shielding'') or from room temperature (``without thermal shielding''). The last buffer of the radiation shield is close to the sample and its temperature is therefore expected to be only slightly higher than that of the sample. In the flow cryostat, the measurements were done with a thermal radiation shield thermally shorted to the sample holder thus having the same temperature as the sample, ensuring a minimum of spurious radiation (``good thermal shielding''). The light source is a commercial infrared LED from SHARP GL4800E0000F with a peak emission wavelength of 0.95~$\mu$m, situated typically 2-3 cm above the sample inside the thermal radiation shield. It is controlled with a Keithley current source 2200. 

The memory effect was measured by cooling down the sample to 2~K under a permanent or shortly applied (typically a few seconds) large dc or ac current (1 to 50~mA). At low temperatures, the excitation was switched off and the resistivity was measured with a low current of 10~$\mu$A upon heating. 

The specific heat was also measured in the PPMS. The magnetization measurements were performed with a SQUID magnetometer from Cryogenics model S700X, the illumination coming from black-body radiation due to the absence of radiation shields. 

To create the mesoscopic samples, we used a Quanta 200 3D dual beam system (ThermoFisher Scientific)---a gallium FIB system combined with a scanning electron microscope---to manufacture micrometric samples. Additionally, the system is equipped with micromanipulators from Kleindiek for in-chamber sample transfer. 
Following a typical transmission electron microscope sample preparation recipe, a Pt protection layer of about 150 nm thickness was deposited by electron beam induced deposition to protect the near surface region of the sample. The thickness of the protection layer was increased to about 2~$\mu$m by an ion beam induced deposition (IBID) of Pt.
The preparation proceeded by removing bulk material around the protected area with the ion beam, ending up in an about 2~$\mu$m thick sample. It was freed from bulk by undercutting, followed by lifting it out, and welding it to the sample grid. In consecutive milling steps with decreasing ion currents (we used 1~nA and 500~pA for the medium thinning and 100~pA for fine milling) the thickness of the sample was reduced to about 300~nm. In order to reduce the FIB induced damage generated by the previous milling steps at 30~kV ion acceleration voltage, a final cleaning of the sample was done with 5~kV and 30~pA.
The grid with the sample was rotated by $90^{\circ}$ and loaded together with the measuring platform to the FIB chamber. After removing the remainder of the protection layer, the sample was welded by ion beam induced tungsten deposition to the micromanipulator tip and cut off from the grid. Subsequently, the sample was transported to the measuring platform, consisting of Ti/Au stripes (5~nm/45~nm thick respectively) on a SiO$_2$/Si substrate, and welded by IBID to the substrate, to stabilize the sample over the electrical contacts. The contacts were fabricated by Pt IBID with Trimethylcyclopentadienyl-platinum [(CH$_3$)$_3$CH$_3$C$_5$H$_4$Pt] as the precursor and the following ion beam parameters: Acceleration voltage 30~kV, beam current 50~pA, dwell time 0.2~$\mu$s, and beam overlap 0~\%. The connection of the sample to the substrate was cut after contact deposition.

The tr-ARPES apparatus consisted of a hemispherical analyzer and a mode-locked Ti:sapphire laser delivering 1.48~eV pump and 5.92~eV probe pulses at the repetition rate of 250~kHz, or at the interval of 4~$\mu$s~\cite{Ishida2014}. The energy and time resolutions were 16~meV and 300~fs, respectively. By utilizing a pin hole attached next to the sample, we estimated the spot diameters of the pump and probe beams to be 250 and 85~$\mu$m, respectively, on the sample surface and also checked that the spot of the pump beam on the sample surface did not shift within 5~$\mu$m when the delay stage in the pump beam path was shifted for 600~ps pump-probe delay. The sample was cleaved by peeling off an adhesive tape attached on the sample in the vacuum chamber at the base pressure of 5$\cdot$10$^{-11}$~Torr. The Fermi energy was carefully determined by standard ARPES (without the pump-pulse procedure, Fig.~\ref{fig10}\textbf{a}) using a Au film evaporated on the sample holder. The probe intensity was reduced until the space-charge-induced shift was less than a few meV so that multi-photon photoelectrons became invisible (see also Ref.~\cite{Yoshikawa2019} for more details on the experimental procedure). Measurements were performed at the temperature of 7~K.}

\noindent \section*{Samples characterization}
\begin{figure}[ht!]
	\centering
		\includegraphics[width=1\textwidth]{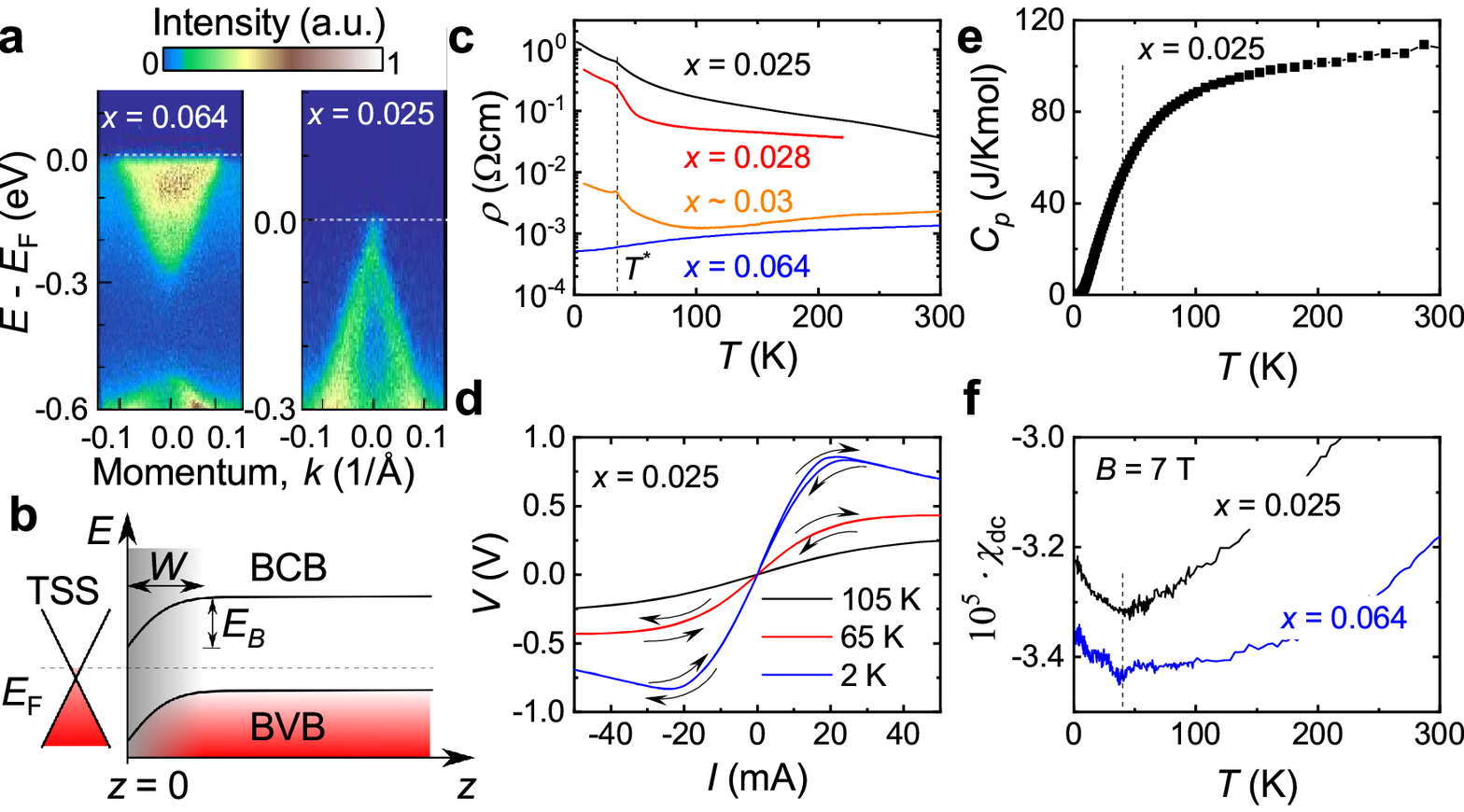}
	\internallinenumbers
	\caption{\textbf{Characterization of Tl$_{1-x}$Bi$_{1+x}$Se$_{2-\delta}$ single crystals.} 
	{\textbf{a}}: ARPES measurement around the $\bar{\Gamma}$ point of the bulk-metallic ($x = 0.064$, left) and bulk-insulating sample ($x = 0.025$, right) at 7\,K. 
	{\textbf{b}}: Sketch of the energy $E$ vs distance from surface $z$ without external stimulus, when the Fermi energy $E_{\text{F}}$ lies in the bulk band gap, showing the formation of a Schottky barrier. $W$ and $E_B$ are the depletion region and barrier height, TSS the topological surface states, and BVB and BCB the bulk valence band the bulk conduction band, respectively.
	{\textbf{c}}: Temperature dependence of the electrical resistivity for samples with different off-stochiometry $x$. An anomaly is visible at $T^* \simeq 40\,$K, independent of $x$. 
	{\textbf{d}}: Current-voltage characteristics of the sample with $x = 0.025$ at selected temperatures without illumination, showing deviations from Ohm's law due to the presence of the Schottky barrier and the development of switching and memory effects below $T^*$. The arrows indicate the measurement sequence: $0 \rightarrow 50\,$mA $\rightarrow -50\,$mA$ \rightarrow 0$. 
	{\textbf{e}}: Temperature dependence of the specific heat in the bulk-insulating sample ($x = 0.025$). 
	{\textbf{f}}: Temperature dependence of the volume magnetic susceptibility $\chi_{\text{dc}}$ in the bulk-insulating ($x = 0.025$, in black) and bulk-metallic ($x = 0.064$, in blue) samples. The measurements were performed at a magnetic field of 7~T to enhance the measurement accuracy, but the same behavior is observed at low fields.
	}
	\label{fig1}
\end{figure}

\MT{Tl$_{1-x}$Bi$_{1+x}$Se$_{2-\delta}$ is a topological insulator featuring} isolated surface Dirac points situated in the band gap (around 300~meV) of the bulk states. The absence of trivial surface states as observed by angle resolved photoemission spectroscopy (ARPES) and a great tunability of the Fermi level by changing the off-stoichiometry via the variation of $x$ \cite{Kuroda2010, Kuroda2015, Eguchi2014} are further properties that make this material particularly suitable for our study. The latter is seen in ARPES as an energy shift of the Fermi level with $x$ and in electrical resistivity measurements as orders of magnitude variation of the low-temperature resistivity with $x$ (Fig.\,\ref{fig1}\textbf{a} and \textbf{c}). The presence of high-mobility electrons (the surface Dirac electrons) and low-mobility (bulk) holes is evidenced from a two-carrier analysis \cite{Eguchi2015, Eguchi2019} (Fig.\,\ref{fig2}).

In the temperature-dependent electrical resistivity curves of samples with $x < 0.064$, we observe semiconducting behavior with an anomaly at a characteristic temperature $T^* \approx 40$~K. As no specific heat anomaly is observed at this temperature (Fig.\,\ref{fig1}\textbf{e}), an underlying thermodynamic phase transition can be ruled out. The dc-susceptibility presents a small up-turn (it increases with decreasing temperature, Fig.\,\ref{fig1}\textbf{f}) below 40~K, which will be discussed later.

In what follows we focus on the bulk-insulating compound $x = 0.025$. By lowering the temperature, its current-voltage characteristics evolve from a regular quasi-linear (ohmic) to a Schottky-like behavior (Figs.\,\ref{fig1}\textbf{d} and \ref{fig3}). This is attributed to the presence of a Schottky barrier between the Dirac surface states and the gapped bulk states due to band bending \cite{Hsieh2009, Checkelsky2011}, as sketched in Fig.\,\ref{fig1}\textbf{b} (see below). Below $T^*$, in addition we observe the appearance of  switching behavior and a memory effect. This anomalous behavior is discussed later and we first describe the photoresponse of the electrical resistivity.

\noindent \section*{Radiation effects} 
Various kinds of (modest) radiation profoundly modify the transport behavior below $T^*$. It is thus important to specify under which condition in terms of shielding the measurements were performed. We will discriminate three different situations: Measurements with the best available thermal radiation shield (``good thermal shielding''), ones with a shortened thermal  radiation shield (leading to black-body radiation from temperatures only slightly above the sample temperature; ``incomplete thermal shielding''), and ones without a thermal radiation shield (leading to black-body radiation from room  temperature; ``without thermal shielding''). In a first set of experiments, we measure the electrical resistivity of the bulk-insulating sample during cooling to 2\,K with good thermal shielding, expose it there to illumination from a near-infrared light emitting diode (NIR LED),
then switch off the LED and measure the electrical resistivity during warming (Fig.\ref{fig4}\textbf{a}). The orange and green curves are for two such runs, with a shorter
and a longer illumination time, respectively, the latter being long enough to reach a steady state
(i.e., even longer irradiation does not lead to a further resistance change). Compared to the measurement without illumination, a
strong decrease of the low-temperature electrical resistivity is observed, up to
an order of magnitude at 2~K, corresponding to a large positive
photoconductivity, with the size of the effect being controlled by the
illumination duration.

Intriguingly, also black-body radiation affects the resistivity, which we
noticed when performing an experiment without proper thermal shielding.
We present two experiments, one with incomplete thermal shielding and one without shielding, as shown in Fig.~\ref{fig4}\textbf{a} and Fig.~\ref{fig5}\textbf{a}. In both cases, the sample was held at 2\,K under
these shielding conditions until a steady state was reached (see Fig.~\ref{fig5}\textbf{b} for the resistance relaxation with incomplete shielding), and then measured
during warming. The experiment with incomplete shielding results in a much weaker resistivity reduction than the experiment without shielding. On the other hand we see that the data from the experiments without thermal shielding and with good thermal shielding but under NIR radiation from the LED are almost identical.

Before proceeding, we note on several points. Firstly, the sensitivity to even spurious thermal radiation (with incomplete thermal shielding) explains the anomalies at $T^*$ in the resistivity
curves taken without intentional illumination (see Fig.\,\ref{fig1}\textbf{c} and Fig.~\ref{fig5}). Secondly, after illumination, the initial high resistance state is recovered by heating the sample above $T^*$. This excludes surface pollution \cite{Hsieh2009, Frantzeskakis2017} and sample aging effects \cite{Taskin2011} as origins because these would not be reversible with temperature. Furthermore the small energy of the involved radiation is unlikely to produce molecular doping such as observed in ARPES \cite{Frantzeskakis2017}. And thirdly, the effect is seen after the
illumination is switched off. Thus, the photoexcited carriers have a lifetime that is, at least, long on the time scale of the resistivity measurement.

To explore the relaxation behavior further, we measured the electrical
resistivity at several fixed temperatures as function of time, both during
illumination with the NIR LED ($0 < t < 7200$~s) and after switching it
off ($t > 7200$\,s, Fig.\,\ref{fig4}\textbf{b}). At high temperatures, the
relaxation is fast on the time scales of our measurements (about 1\,s) but with decreasing
temperature an increasingly slow relaxation is observed (spurious heating
effects are addressed in Fig.~\ref{fig6}). Most
striking is the effect that when switching the LED off after having
reached the equilibrium state at 10\,K, the resistivity barely evolves on the
time scale of our experiment (see lowest temperature curve in
Fig.\,\ref{fig4}\textbf{b}). This is the phenomenon of persistent photoconductivity.

We also analyze the time dependence quantitatively. The relaxation after
switching the LED off follows a stretched-exponential relaxation
\cite{Phillips1996}, $R\propto \exp{[-(\frac{t}{\tau})^{n}]}$ (with $n<1$), as shown by the black lines in Fig.\,\ref{fig4}\textbf{b} (with the condition $R_{t \rightarrow \infty} = R_{t<0}$ during relaxation). The temperature
dependence of the relaxation time $\tau$ is shown in Fig.\,\ref{fig4}\textbf{c}. It increases by almost 9 orders of magnitude as temperature is only slightly reduced, from 35 to 20~K, before saturating at lower temperatures. 
Above 20~K, this temperature dependence can be approximated by a thermal activation law, $\ln{(\tau)} \propto E_B/k_{\text{B}}T$, with $E_B \approx 82$\,meV interpreted as the height of a Schottky barrier.
 From this dependence we derive that, below 20\,K, a relaxation of the resistivity to 90\% of its initial value before irradiation would take hundreds of years---persistence par excellence (see Fig.~\ref{fig6}).

\noindent \section*{Current effects} 
Next we show that the surface conductivity can also be tuned using electrical currents.
Applying a large current at low temperatures leads to a controlled and permanent decrease of the resistivity which, analogous to the illumination experiments, is measured upon warming, after reducing the large tuning current to a small measurement current of 10~$\mu$A to exclude overheating (Fig.\,\ref{fig4}\textbf{d}). Again, the effect is reversible by heating the sample above $T^*$.

This current-induced decrease of the resistivity below $T^*$ is visible in the current-voltage characteristics as a
memory effect. In the 2\,K isotherm (Fig.\,\ref{fig1}\textbf{d}), a hysteresis is observed between the increasing-current curve (from 0 to 50\,mA) and the decreasing-current curve (from 50\,mA back to 0) but, interestingly, not between the subsequent up and down sweeps with negative applied current. The hysteresis can thus be attributed to the current-induced switching from the initial high-resistance state (black line in Fig.\,\ref{fig4}\textbf{d}) to the low-resistance state (purple triangles). Once the sample has switched, it remains in the low-resistance state until it is heated above  $T^*$. Thus, no hysteresis appears under negative currents, and not under positive currents in a second loop either. The characteristic time of the current-induced resistance change is much faster than that of the illumination-induced process; in fact, it is smaller than the typical time scale of our measurements (about 1\,s) and could thus not be resolved.

\MT{To understand the Schottky-like behavior, we model the topological insulator as a metal-semiconductor-metal device, the ``metal'' being the surface and the ``semiconductor'' the bulk. Charge injection from the metal to the semiconductor is frequently modeled by the relation\cite{Sze2007} $I=I_0(e^{\frac{qV}{\eta k_{\text{B}} T}}-1)$, which captures both thermionic emission and tunneling processes. Here $I_0$ is the saturation current, $q$ the elementary charge, $V$ the applied voltage, $k_{\text{B}}$ the Boltzmann constant, $T$ the temperature, and $\eta$ an ideality factor\cite{Sze2007}. The ohmic regime at low current is taken into account by adding a linear term, giving $I=I_0(e^{\frac{qV}{\eta k_{\text{B}} T}}-1)+V/R$, where $R$ is the resistance of the sample. With $q=+e$ (the majority carriers being holes), our current-voltage characteristics can be rather well described by the fit, as seen by the red lines in Fig.~\ref{fig3}\textbf{a}. The activation energy $E_{\text{A}}$ can be obtained through $I_0$ as $\text{ln}(I_0/T^2) \propto E_{\text{A}}/k_{\text{B}} T$ or through $\ln{R} \propto E_{\text{A}}/2k_{\text{B}}T$ [\citenum{Sze2007}]. The fact that one obtains similar activation energies in both cases (28 and 25~meV, respectively, see the blue lines in Fig.~\ref{fig3}\textbf{b}) indicates that the model captures the situation reasonable well. The ratio of tunneling to thermionic components is given by the ideality factor $\eta$. Departure from unity indicates that tunneling becomes the dominant process, as observed in our case when lowering the temperature, as shown in Fig.~\ref{fig3}\textbf{c}.
At high currents, charge injection directly into the conduction band occurs, giving an increase of the measured voltage upon illumination (shown by the arrows in Fig.~\ref{fig3}\textbf{d}). At low temperatures, the effect of the illumination changes at the threshold current: it is reduced at small currents, and enhanced at large currents. This indicates that the switching is caused by carrier injection into the conduction band.}

\noindent \section*{Discussion}
How can we understand the salient features observed in all these experiments in a consistent way? From the similarities between the light-induced and electrical current-driven reduction of the resistivity, a common mechanism is expected and therefore both purely optical \cite{Yeats2015, Pirralho2019} and purely electric-field-induced effects\cite{Fu2013, Kalkan2015, Tian2017} are excluded. In addition, we observe essentially the same behavior in samples with different off-stoichiometry/doping level, in microfabricated samples (see below), and in a bulk metallic sample (when the Fermi level is in the bulk conduction band), see Fig.~\ref{fig7}. This indicates that the effects are robust against changes of the Fermi level, impurity concentrations, and eventual surface potential fluctuations, therefore discarding trapped charges in the band gap as origin \cite{Lang1977,Zukotynski1987,Scalvi2020}. Finally, both freshly cleaved and ``older'' samples behave similarly, with a reversible behavior under excitation (heating brings back the original state). This points to an intrinsic mechanism, with only little to no influence from surface oxidation. It also suggests that the topological states are directly involved, as no topologically trivial surface states have been observed in ARPES (on cleaved surfaces) in TlBiSe$_2$ , see e.g. \cite{Kuroda2010,Pielmeier2015,Singh2016}.

We first address the effects induced by the NIR radiation, i.e., the large positive and persistent photoconductivity with stretched-exponential relaxation, beginning with the determination of the volume of the sample affected by illumination. The depletion depth $W$ (see Fig.\,\ref{fig1}) is estimated by solving the Poisson equation \cite{Sze2007}, giving $W=\sqrt{\frac{2\epsilon_S}{e^2n}(E_B - eV - k_{\text{B}}T)}$, where $\epsilon_S$ is the permittivity of the material, $e$ the elementary charge, $n$ the carrier density, $E_B$ the barrier height, $V$ the applied voltage, $k_{\text{B}}$ the Boltzmann constant, and $T$ the temperature. Using $n =10^{22}$~m$^{-3}$, the permittivity $\epsilon_S=21\epsilon_0$ [\citenum{Konorov2006}], and $E_B=82$~meV obtained from the temperature dependence of the relaxation time (Fig.~\ref{fig4}\textbf{c}), one finds $W \approx 15$~nm at low temperatures without applied voltage. An estimate of the light penetration depths using the light source is obtained using early studies of TlBiSe$_2$ thin films\cite{Mitsas1992, Mitsas1993}: Depending on the growth condition, the light penetration depth at a wavelength $\lambda=2$~$\mu$m (optical response around 1~$\mu$m not reported to the best of our knowledge) varies between 0.16 and 1~$\mu$m. 
The penetration depth of the NIR light is thus much larger than the depletion depth $W$, and has sufficient energy ($\approx 1.25$\,eV) to excite carriers across the bulk bandgap ($\ge 0.3$\,eV \cite{Kuroda2010}). Thus, we propose that electrons, photoexcited across the bulk bandgap, migrate to the surface due to the downward band bending associated with a Schottky barrier,
and populate the Dirac surface states (as sketched in Fig.\,\ref{fig4}\textbf{e}\,top) at the crossover temperature $T^*$ (for instance when the generation rate becomes larger than the recombination rate as suggested in Ref.~\cite{Pirralho2019}). This creates a surface charge accumulation with space-charge separation\cite{Queisser1985, Queisser1986} between the Dirac electrons and the bulk holes. By populating the Dirac states, the barrier height $E_B$ and, as a consequence, the depletion width $W$ decrease. 

\begin{figure}[ht!]
	\centering
		\includegraphics[width=1\textwidth]{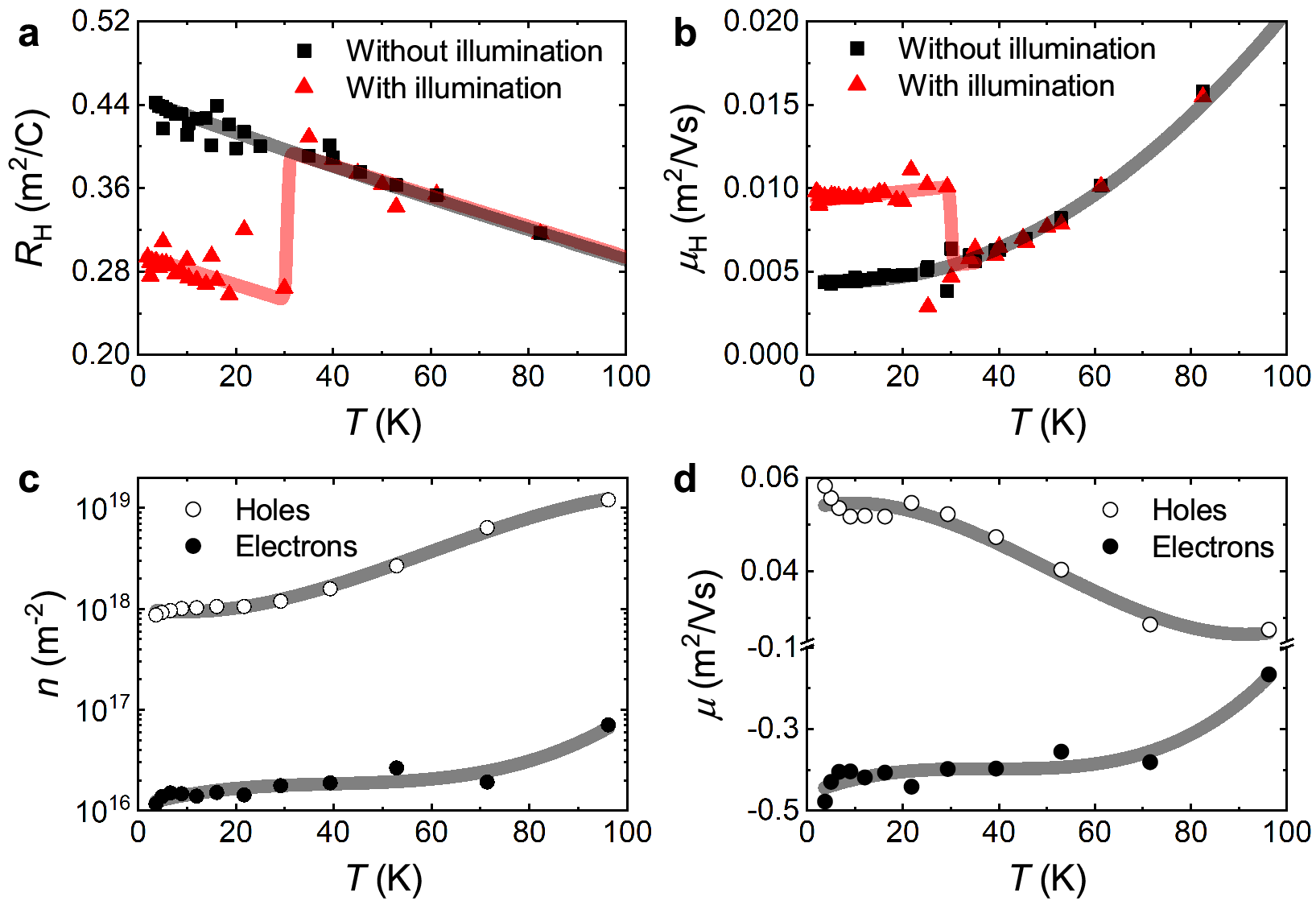}
	\internallinenumbers
	\caption{\textbf{Photoresponse of the Hall coefficient and the Hall mobility.} \textbf{a}, \textbf{b}: Linear-response Hall coefficient $R_{\text{H}}$ and Hall mobility $\mu_{\text{H}}$ without and with illumination (black and red symbols, respectively). Without external excitation, the sheet carrier concentrations and mobilities of the bulk holes and Dirac electrons, estimated using a two-carrier analysis, are given in panels \textbf{c} and \textbf{d}, respectively. The thick lines are guides to the eyes.
	}
	\label{fig2}
\end{figure}

\begin{figure}[ht!]
	\centering
		\includegraphics[width=1\textwidth]{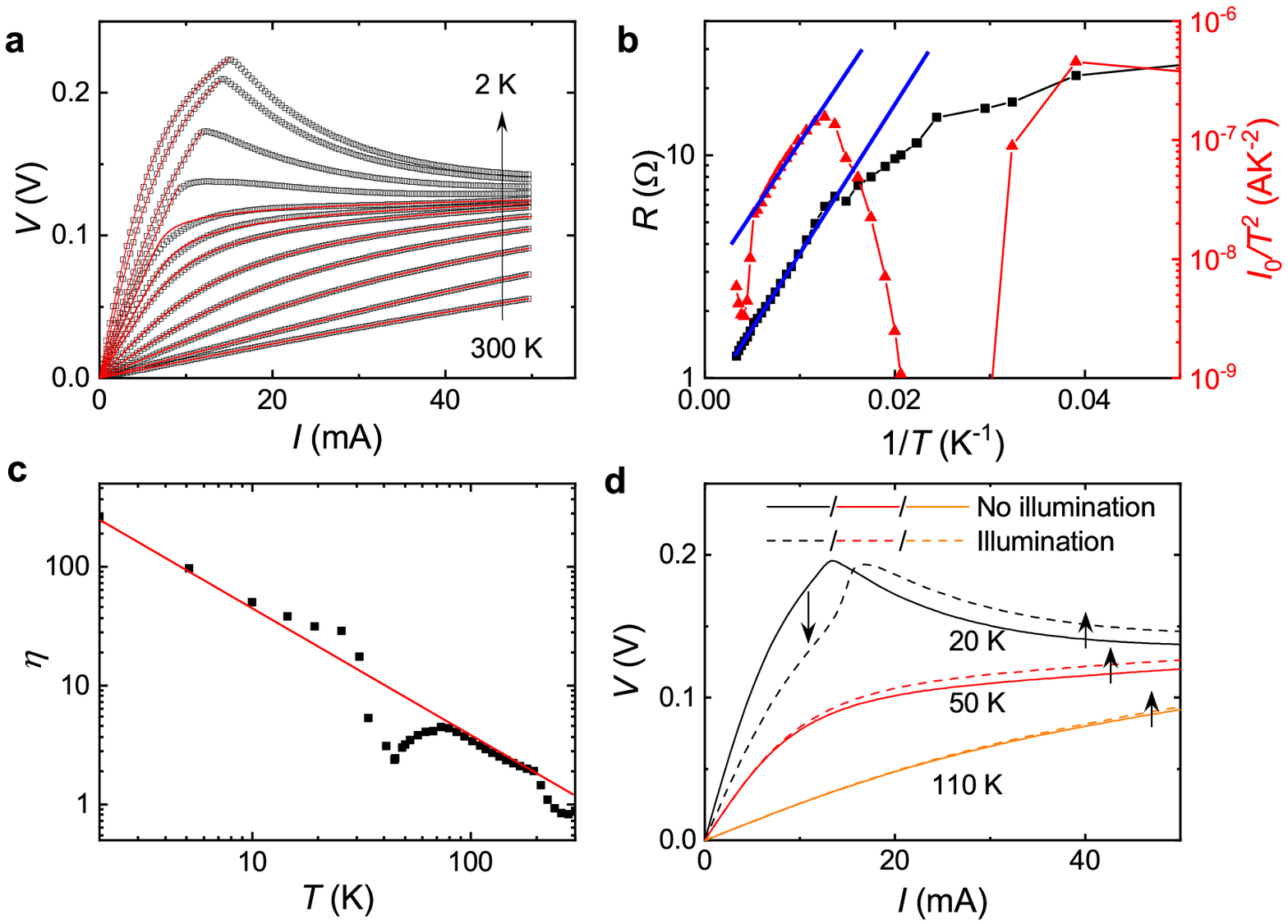}
	\internallinenumbers
	\caption{\textbf{Current-voltage characteristics.} \textbf{a}: Temperature dependence of current-voltage characteristics without illumination; they can be described as a metal-semiconductor junction (red lines, see text). \textbf{b}: Resistance (in black, left axis) and $I_0/T^2$ (red, right axis, see text) versus $1/T$; the blue lines highlight the range of thermally activated behavior. \textbf{c}: Temperature dependence of the ideality factor $\eta$; the red line is a guide to the eyes. \textbf{d}: Photoresponse of the current-voltage characteristics, shown by the arrows. For all measurements, a current of 50~mA was briefly applied to switch the sample into the ``low resistance'' state (purple triangle of Fig.~\ref{fig4}\textbf{d}). For temperatures above $T^*$, the characteristics present an increase of voltage under illumination in the whole current range. Below $T^*$, the voltage is reduced upon illumination at small currents, but enhanced at larger currents, after the switching.
	}
	\label{fig3}
\end{figure}

\begin{figure}[ht!]
	\centering
		\includegraphics[width=1\textwidth]{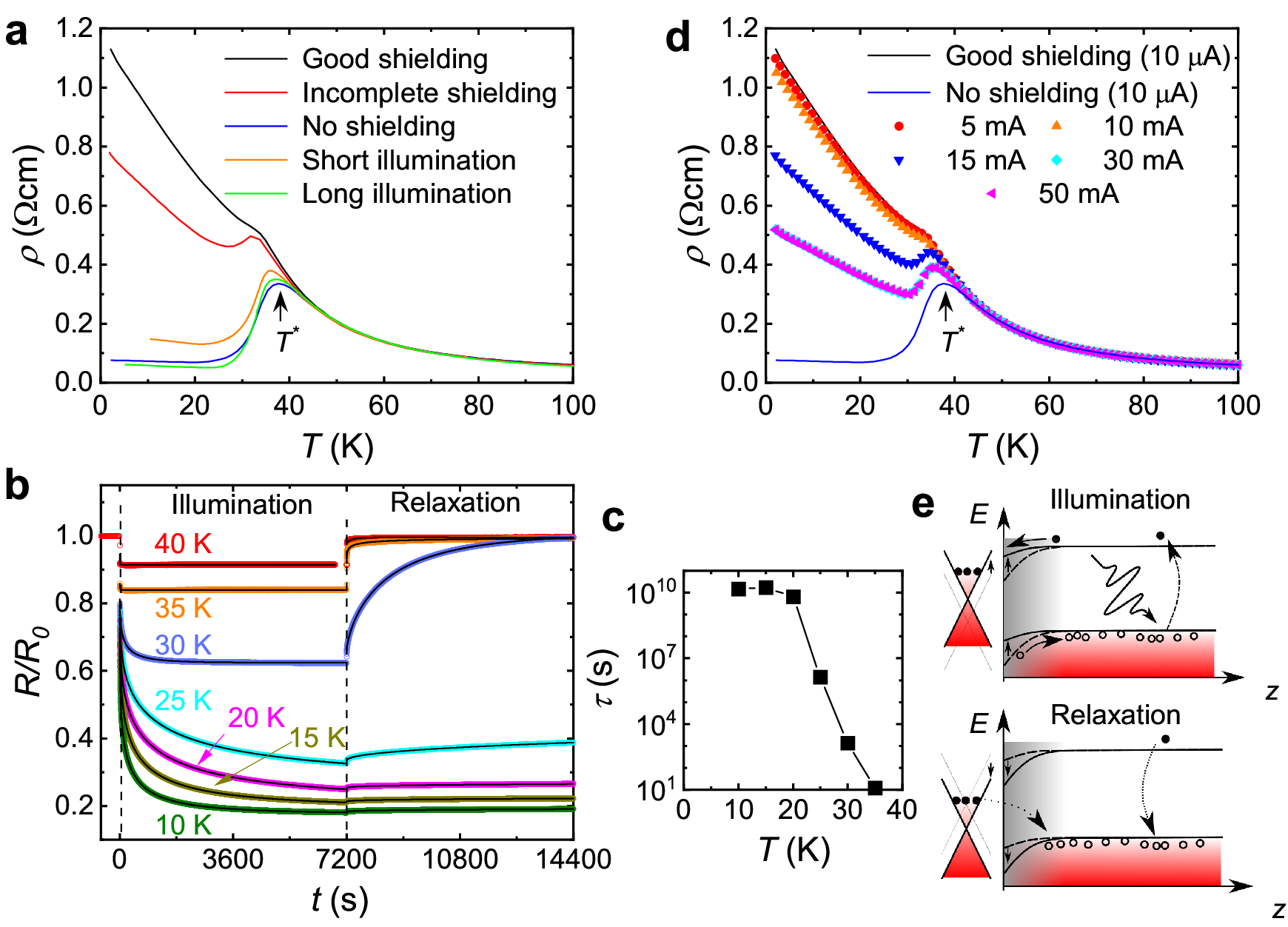}
	\internallinenumbers
	\caption{\textbf{Tuning the surface conductivity of Tl$_{1-x}$Bi$_{1+x}$Se$_{2-\delta}$ ($x = 0.025$) in transport.} {\textbf{a}}: Photocontrol of the resistivity under different conditions. The poorer the thermal shielding (black $\rightarrow$ red $\rightarrow$ blue), the stronger the resistivity suppression below $T^*$ (see also Fig.~\ref{fig5}). Intentional illumination with a NIR source (orange, green) was applied with two different exposure times at 2\,K, and switched off just before the measurements were performed upon warming. All measurements were done with an excitation current of $10\,\mu$A.
	\textbf{b}: Time dependence of the normalized resistance upon illumination below 40~K. The illumination is on between $t=0$ and 7200~s, and then is switched off. The black solid lines are fits using a stretched-exponential (see text). \textbf{c}: Relaxation time $\tau$ as function of temperature, reaching 317 years at 10\,K.
	{\textbf{d}}: Current-control of the resistivity, with different currents applied at 2\,K before the measurements, which were then performed with $10\,\mu$A upon warming. All current-controlled measurements 
	were performed under good thermal shielding. The blue and black curves are reproduced from {\textbf{a}} for comparison.
   \textbf{e}: Energy vs distance from surface diagrams, sketching the illumination and relaxation process. Under illumination, photoexcited electrons and holes are generated in the bulk, and migrate to the surface and bulk states, respectively, due to the downward band-bending. This leads to surface charge accumulation and space-charge separation, modifying thus the band-bending and shifting the Dirac cone accordingly. After turning off the illumination, the excited Dirac electrons recombine with the excited bulk holes, which at low temperatures can only occur via tunneling.
	}
	\label{fig4}
\end{figure}

\begin{figure}[ht!]
	\centering
		\includegraphics[width=1\textwidth]{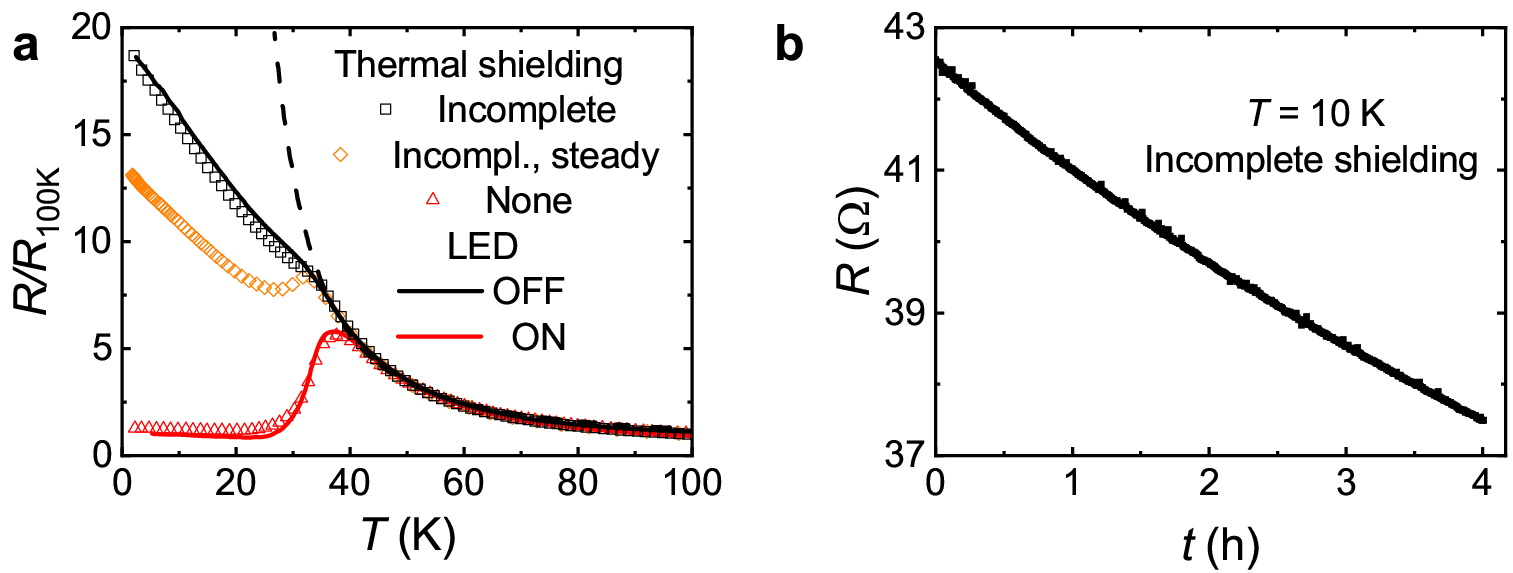}
	\internallinenumbers
	\caption{\textbf{Comparison of illumination with infrared light and with black-body radiation.} \textbf{a}: The measurements with the infrared source (LED) and under black-body radiation from room temperature (without thermal shielding) give similar results (in red). Whereas with the good thermal shielding, the anomaly at $T^*$ is barely visible (black line), a small anomaly appears with incomplete thermal shielding (black points). In this case, the resistance below $T^*$ slowly drifts to lower values (orange points) as shown in panel \textbf{b}. The dashed line in panel a is the extrapolation of the Arrhenius fit to the resistivity data above 40~K to temperatures below 40~K.
	}
	\label{fig5}
\end{figure}

\begin{figure}[ht!]
	\centering
		\includegraphics[width=0.5\textwidth]{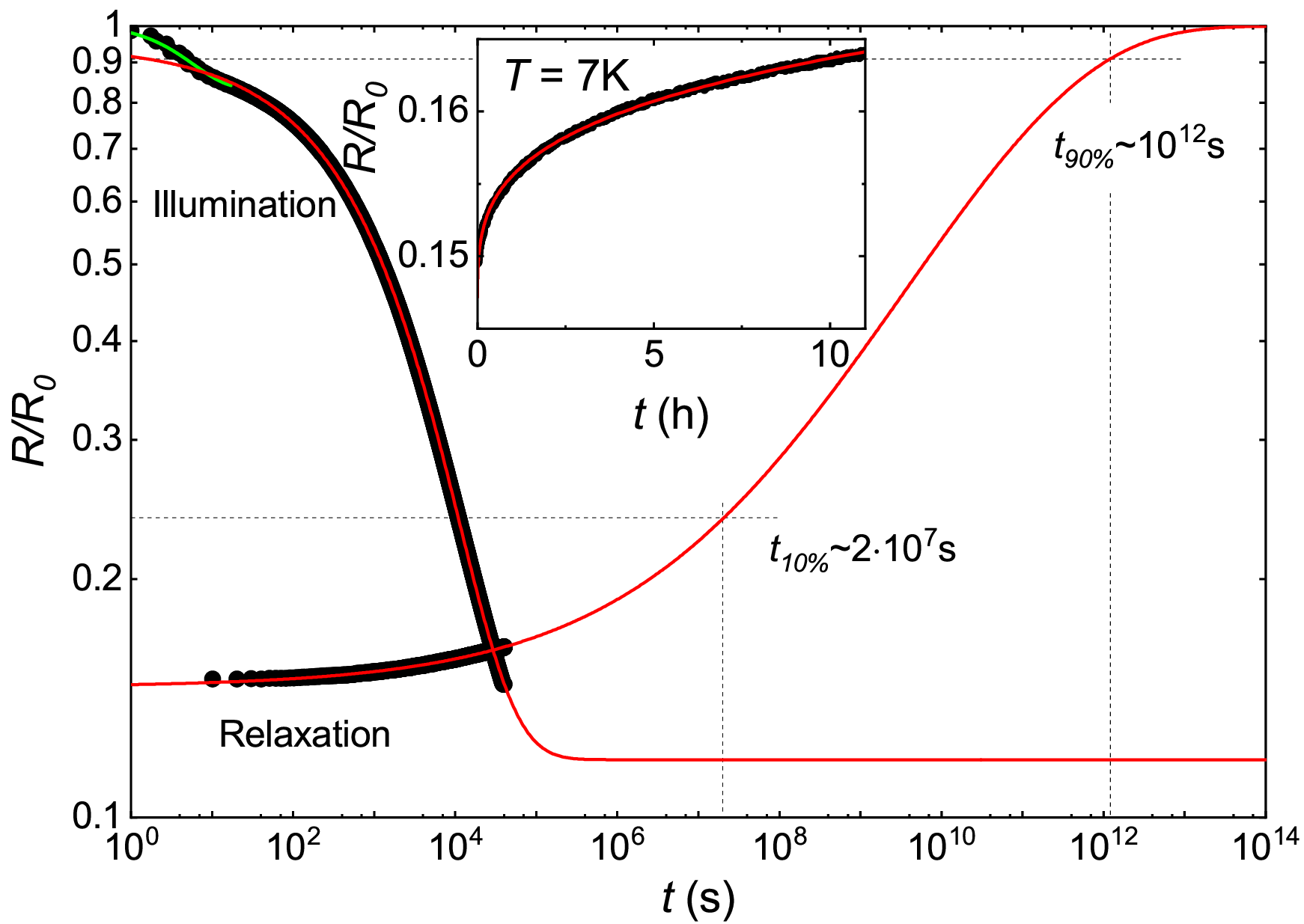}
	\internallinenumbers
	\caption{\textbf{Details on the illumination/relaxation process at 7~K.} Normalized resistance during illumination and relaxation (black points). The red lines correspond to fits using stretched exponentials, with the constraint $R_{t\rightarrow\infty}=R_0$ for the relaxation process. 
	 During illumination, an initial fast exponential decay is observed (green line); we attribute it to overheating when the LED is switched on. Insert: Relaxation process in linear scale with the fit. During relaxation, in 11~hours, the resistance has recovered less than 2\% of its initial value.
	}
	\label{fig6}
\end{figure}

\begin{figure}[ht!]
	\centering
		\includegraphics[width=0.5\textwidth]{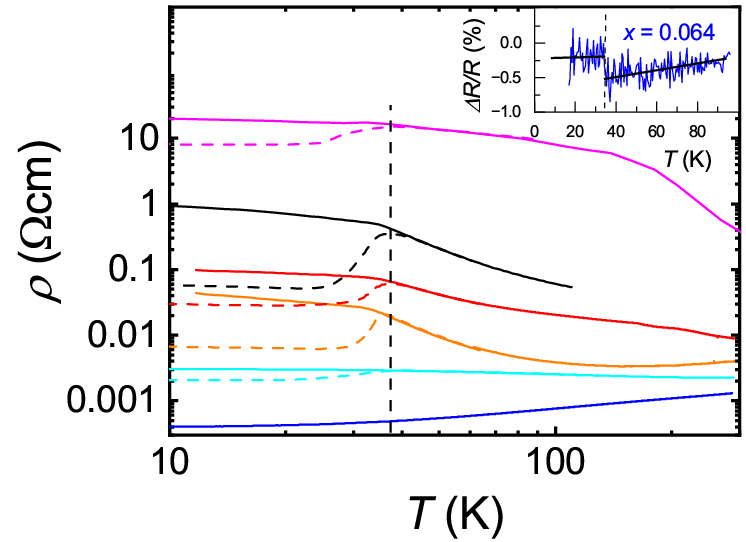}
	\caption{\textbf{Effect of the illumination on different samples.} Resistivity measurements in samples with different off-stoichiometry/doping level, showing the transition from insulating to metallic behavior, without illumination (solid lines) and under illumination (dashed lines). In the case of the bulk metallic sample ($x = 0.064$, in blue), one can still see an anomaly at $T^*$ by looking at the difference of the resistance with and without illumination (insert, note that in this case the illumination is kept on during measurements). Here, one defines $\Delta R / R = (R_{off}-R_{on})/R_{off}$, with $R_{off}$ and $R_{on}$ the resistance without and with illumination respectively. The solid black lines are guides to the eyes.
	}
	\label{fig7}
\end{figure}

This model is supported by Hall effect experiments. They reveal a pronounced illumination-induced change of the Hall coefficient below $T^*$ (Fig.~\,\ref{fig2})indicating that a new conduction channel opens, namely the region of the sample into which the light penetrates. Thus, a simple two-carrier model \cite{Eguchi2019} is not applicable and an extension is needed. With the following reasonable extra assumptions one can nevertheless extract useful information. We assume that the same amount of electrons and holes is generated by the illumination, that the excited electrons migrate to the surface states (where they have the same mobility as without illumination), and that the excited holes represent the new channel and are thus allowed to have a mobility different from the bulk holes. The observed decrease of the (positive) Hall coefficient and increase of the Hall mobility (Fig.~\ref{fig2}\textbf{a}, \textbf{b}) are then only feasible if these extra holes exhibit an increased mobility.
With these two assumptions, we estimate a surface carrier generation of around $1 \cdot 10^{17}$~m$^{-2}$ at low temperature under illuminations (8 times the surface carrier concentration without illumination) and an increase of the hole mobility by a factor $3.5$. To estimate the weight of the excited electrons/bulk holes in the total conductivity, one assumes that the conductivity of each carrier is given by $\sigma_i = q n_i \mu_i$, with $q$ the elementary charge of the charge carrier $i$ (electron or hole), and $n_i$ and $\mu_i$ the corresponding carrier concentration and mobility respectively. At 10~K, it is $\sigma_e/(\sigma_h+\sigma_e) \approx 10$~\% without illumination, and $\sigma_e/(\sigma_h+\sigma_e) \approx 24$~\% under illumination, an increase by a factor 2.4.

The stretched-exponential decay is understood as follows: the photoexcited Dirac
electrons on the surface are separated from the photoexcited holes in the bulk
by the Schottky barrier. At low temperatures, the recombination process is
dominated by tunneling across the depleted area (as sketched in
Fig.\,\ref{fig4}\textbf{e}\,bottom). The barrier height $E_B$ and depletion width $W$ increase as the Dirac states are depopulated. As a consequence, the tunneling rate decreases with time during relaxation, compatible with the observed stretched-exponential decay.

We now turn to the sensitivity to spurious black-body radiation (without thermal shielding). The power density from a black body at 300\,K is centered at 155\,meV
(8\,$\mu$m) with a significant part of the distribution above 300~meV, which is large enough to excite carriers across the energy gap. The steady-state photoconductivity from this radiation is almost identical to the one from the much higher energy NIR LED (see Fig.~\ref{fig4}\textbf{a}). This shows that, as long as the radiation has sufficient energy to excite carriers across the energy gap, a unique steady-state conductivity is reached. This provides further support for the conductivity in the irradiated state being dominated by surface conduction because the different energy ranges of the two types of radiations imply different penetration depths into the sample. If this resistivity reduction were due to (metastable) bulk carriers, different portions of the bulk sample would contribute and thus different saturation resistivities would be observed. By contrast, black-body radiation from much lower temperature in our experiments with incomplete thermal shielding has insufficient energy to excite carriers across the bulk gap. However, because bulk-insulating Tl$_{1-x}$Bi$_{1+x}$Se$_{2-\delta}$ is slightly off-stoichiometric, there may be shallow in-gap states with trapped carriers, and these can still be photoexcited to the bulk conduction (or valence) band. This can explain the weak resistivity reduction of the curve with incomplete shielding (red curve in Fig.\,\ref{fig4}\textbf{a}), the smaller anomaly in the resistivity (as shown by the vertical dashed line in Fig.~\ref{fig1}\textbf{c}), and the still sizable deviation from the Arrhenius behavior below 40~K (dashed line in Fig.~\ref{fig5}). Most likely, such unintended spurious radiation effects are responsible for the unexpected ultra-long/persistent behavior observed in Refs.~\cite{Tian2017,Li2017,Uchiyama2023}. Furthermore, the associated charge accumulation adds a Pauli paramagnetic contribution to the diamagnetic dc susceptibility, which leads to the up-turn in the dc susceptibility mentioned above (Fig.~\ref{fig1}\textbf{f}).

Next we discuss the electrical current-driven effect. As shown above, there is a current threshold (between 10 and 15\,mA, Figs.\,\ref{fig1}\textbf{d} and \ref{fig3}) for low-temperature resistance switching to occur. For currents (and corresponding electrical fields) above this threshold, the bulk conduction band is partially populated by electrons via electric field injection \cite{Sze2007}. At low temperatures, these electrons then increase the ``current-induced conductivity'', analogous to the photoconductivity under irradiation.

\noindent \section*{Additional transport results}
\begin{figure}[ht!]
	\centering
		\includegraphics[width=1\textwidth]{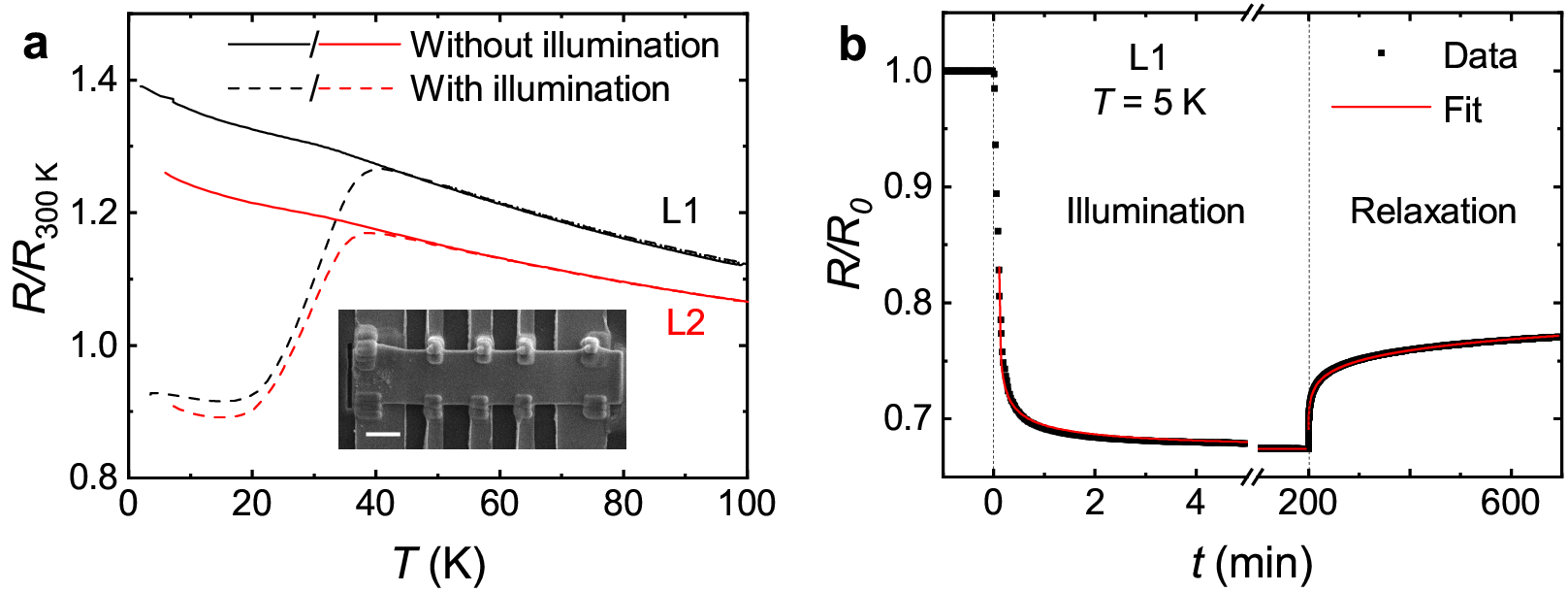}
	\internallinenumbers
	\caption{\textbf{Photoconductivity in micro-fabricated Tl$_{1-x}$Bi$_{1+x}$Se$_{2-\delta}$ ($x = 0.025$) samples with good thermal shielding.} \textbf{a}: Temperature dependence of the resistance, normalized to its value at room temperature, without and with illumination in two samples (L1 and L2). In inset is a scanning electron microscopy image of a typical structure, the scale bar corresponds to 5 $\mu$m. \textbf{b}: Time dependence upon illumination at 5~K in sample L1. A stretched-exponential relaxation is observed (red line), with an ultralong time constant of $9 \cdot 10^6\,$s estimated
 for the relaxation process.
	}
	\label{fig8}
\end{figure}

We now present photocontrol studies on micrometric samples, fabricated with the focused-ion beam (FIB) cutting technique \cite{Friedensen2017} from a bulk sample ($x = 0.025$) (insert of Fig.\,\ref{fig8}\textbf{a}). The temperature-dependent resistances of two such samples again show a pronounced drop below about 40\,K when illuminated and the time dependence upon illumination again follows stretched-exponential behavior (Fig.\,\ref{fig8}\textbf{a} and \textbf{b}). Both this qualitative similarity to the results on the bulk sample and two quantitative differences, reported next, lend further support to our interpretation, and confirm that the surface states are directly involved. Firstly, during illumination, the saturation resistance is reached in only a few minutes instead of a few hours, which we attribute to the light penetration depth being similar to or larger than the thickness of the sample ($0.16-1\,\mu$m and 0.3~$\mu$m, respectively), and thus the irradiation hitting the entire volume of the sample. Secondly, the resistance under illumination is lower at low temperature than it is at room temperature (ratio $R/R_{\text{300~K}} < 1$), which we understand as caused by a much larger relative contribution of the surface conduction in the samples. Our results demonstrate the robustness of the photocontrol of Dirac surface conduction against microfabrication, thus paving the way for device applications.

\noindent \section*{ARPES} 
\begin{figure}[ht!]
	\centering
		\includegraphics[width=1\textwidth]{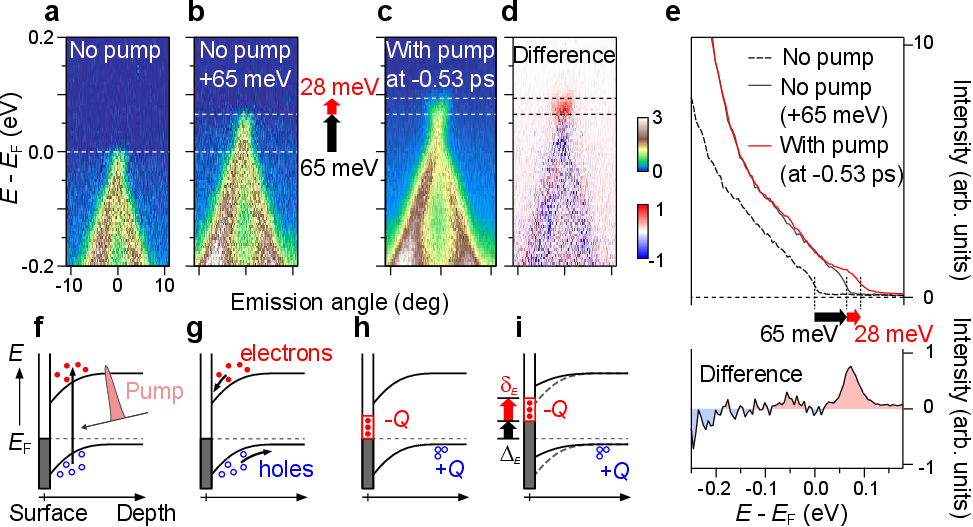}
	\internallinenumbers
	\caption{{\bf Tr-ARPES in  Tl$_{1-x}$Bi$_{1+x}$Se$_{2-\delta}$ ($x = 0.025$) at 7~K.}  {\bf a-c}: Dispersion of the surface states around the $\bar{\Gamma}$ point without pump (\textbf{a}) and after the longest possible pump-probe delay of 4~$\mu$s (\textbf{c}). The spectrum in \textbf{b} corresponds to the one in \textbf{a} shifted by $+65$\,meV. {\bf d}: Difference between the distributions shown in {\bf b} and {\bf c}. {\bf e}: Angle-integrated energy distribution curves of the spectra shown in {\bf a}-{\bf c} (top), and difference between the red and solid black curves (bottom). The interval of integration was [-14.5$^{\circ}$, 14.5$^{\circ}$]. {\bf f}-{\bf i}: Schematic of the photovoltage generated on the surface of a bulk-insulating topological insulator (see text).
	}
	\label{fig9}
\end{figure}

\begin{figure}[ht!]
	\centering
		\includegraphics[width=1\textwidth]{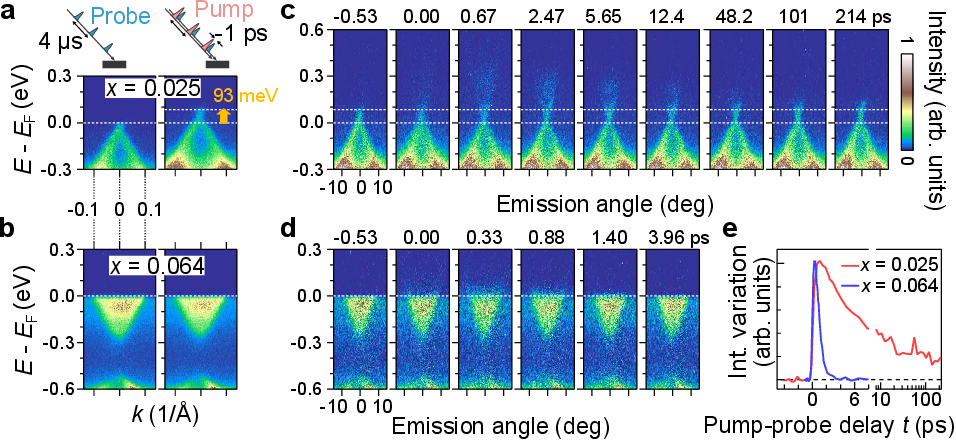}
	\internallinenumbers
	\caption{\textbf{Tr-ARPES for several pump-probe delays at 7~K.} \textbf{a}: Spectra around the $\bar{\Gamma}$ point in the bulk-insulating sample ($x=0.025$) without (left) and with (right) a pump, at the longest possible pump-probe delay of 4$\mu$s, where a surface shift of 93~meV is still present. On top is a sketch of the pump-probe procedure. \textbf{b}: Same as \textbf{a} in the bulk-metallic sample ($x=0.064$). \textbf{c}: Tr-ARPES spectra at several pump-probe delays (in ps) in the bulk-insulating sample ($x=0.025$). The negative delay is equivalent to a pump-probe delay slightly shorter than 4~$\mu$s. \textbf{d}: Same as \textbf{c} in the bulk-metallic sample ($x=0.064$). \textbf{e}: Variation of the intensity as a function of the pump-probe delay in both samples, obtained by subtracting the intensity of the spectrum at 4~$\mu$s (average of 10 images) from the one of each spectrum at various pump-probe delays. The interval of the integration was over the angle [-14.5$^{\circ}$, 14.5$^{\circ}$] and the energy $E>0$ (states above the Fermi level).
	}
	\label{fig10}
\end{figure}

Finally we studied the Dirac surface states with time-resolved ARPES (tr-ARPES) experiments. In the bulk-insulating sample ($x=0.025$), after pumping, a shift of the (quasi-)Fermi level by 93\,meV is observed and persists for at least $4\,\mu$s (Fig.~\ref{fig9}\textbf{a},\,\textbf{c} and Fig.~\ref{fig10}\textbf{a},\,\textbf{c}). This shift is composed of an actual shift of the Fermi level on the surface by $\delta_E=28$\,meV---caused by the filling of the Dirac states upon pumping (Fig.~\ref{fig9}\textbf{f-h})---and of a shift by $\varDelta_E=65$\,meV caused by a surface photovoltage due to a relaxation of the band bending (Fig.~\ref{fig9}\textbf{g}). To visualize the charge accumulation at the surface more directly, we subtract the spectrum without pump but shifted by 65\,meV (Fig.~\ref{fig9}\textbf{b}) from the spectrum taken 4~$\mu$s after the pump (Fig.~\ref{fig9}\textbf{c}). The difference spectrum (Fig.~\ref{fig9}\textbf{d}) as well as the difference of the angle-integrated intensity (Fig.~\ref{fig9}\textbf{e}) clearly reveal extra charge corresponding to a Fermi level shift of 28\,meV. No such ``persistent'' shift of the Fermi level is detected in the bulk-metallic sample (Fig.~\ref{fig10}\textbf{b}, \textbf{d}). In addition, very different relaxation times of the excited states between both samples are found (Fig.~\ref{fig10}\textbf{e}). Only if the Fermi level is situated within the bulk band gap, we observe a slow ($\ge 4\,\mu$s) relaxation, similar to that observed in other topological insulators\cite{Neupane2015, Sumida2017, Yoshikawa2018, Papalazarou2018, Yoshikawa2019, Ciocys2020}. Contrary to those studies which focused only on spectroscopic probes, our accompanying transport study on the same material allows us to relate this ``more-than-$4\,\mu$s-long'' relaxation to an ultraslow ``more-than-hour-long'' relaxation. We conjecture that the underlying mechanism is the same and that, thus, also the tr-ARPES pump induces a state of persistent topological surface conduction in bulk insulating Tl$_{1-x}$Bi$_{1+x}$Se$_{2-\delta}$ ($x=0.025$).

\noindent \section*{Summary and conclusion}
We have demonstrated in this work that topological surface states can be permanently populated by carriers, enhancing the conductivity of the surface states. This is achieved using simple external stimuli, notably illumination and electrical current. The mechanism proposed here---space-charge separation between the surface and bulk carriers---is relevant for any topological insulator. It is surprising that these effects have remained largely unnoticed in transport and poorly understood \cite{Yeats2015,Zheng2015,Tian2017,Li2017,Uchiyama2023,Pirralho2019,Kang2019,Hou2020,Xie2021}. Possibly, the anomaly in transport observed in Cd-doped Bi$_2$Se$_3$\cite{Ren2011} and Sn-doped Bi$_{1.1}$Sb$_{0.9}$Te$_2$S \cite{Kushwaha2016} are analogous to what we observe here, just as the large difference between (dc-)resistivity and optical conductivity (therefore after irradiation) in BiSbTeSe$_2$\cite{Borgwardt2016} and the persistence of the voltage-induced superconductivity in Bi$_2$Se$_3$\cite{Le2021}.

We expect our work to have broad impact. Firstly, our demonstration of the extreme sensitivity of the surface states to even weak irradiation will raise awareness that extra care needs to be taken in order to probe the ground state of a topological insulator, particularly in spectroscopic experiments where shielding is difficult.
Secondly, we have put forward experiments with readily accessible ``pumping''---weak light/radiation or electrical current---which lend themselves for experiments also on other samples in many laboratories worldwide, as well as for implementation in experiments beyond transport. An interesting candidate for such future studies is the putative topological Kondo insulator SmB$_6$\cite{Ishida2015, Pirie2021}. 
 Thirdly, because of the persistence of the surface conduction, the ``pump'' can be switched off after the surface charge is generated. This will avoid any interference with it both in fundamental studies and future device applications.
And finally, the control does not rely on the presence of defect or impurity states and is even robust against rather harsh microstructuring techniques, ensuring reliability. 


\section*{Acknowledgements}
We thank Y.~Ando, A.~V.~Balatsky, F.~Libisch, S.~Rotter, and M.~Shiraishi for discussion, J.~Baraillon and P.~Hofegger for technical assistance during the measurements, and M.~Schinnerl and W.~Schrenk for assistance in the cleanroom. Part of this work was done in the cleanroom facilities ZMNS of TU Wien. We acknowledge financial support by the Austrian Science Fund (FWF Grants No. 29279-N27 and No. SFB F 86), the European Research Council (ERC Advanced Grant 101055088) and by KAKENHI (18H01148, 17H06138, 18H03683).
\newpage



\begin{thebibliography}{10}
\expandafter\ifx\csname url\endcsname\relax
  \def\url#1{\texttt{#1}}\fi
\expandafter\ifx\csname urlprefix\endcsname\relax\def\urlprefix{URL }\fi
\providecommand{\bibinfo}[2]{#2}
\providecommand{\eprint}[2][]{\url{#2}}

\bibitem{Hasan2010}
\bibinfo{author}{M.~Z. Hasan}, \ and\ \bibinfo{author}{C.~L. Kane},
\newblock \bibinfo{title}{{Colloquium: Topological insulators}}.
\newblock \emph{\bibinfo{journal}{Rev. Mod. Phys.}}
  \textbf{\bibinfo{volume}{82}}, \bibinfo{pages}{3045}
  (\bibinfo{year}{2010}).

\bibitem{Qi2011}
\bibinfo{author}{X.-L. Qi}, \ and\ \bibinfo{author}{S.-C. Zhang},
\newblock \bibinfo{title}{{Topological insulators and superconductors}}.
\newblock \emph{\bibinfo{journal}{Rev. Mod. Phys.}}
  \textbf{\bibinfo{volume}{83}}, \bibinfo{pages}{1057}
  (\bibinfo{year}{2011}).

\bibitem{Wehling2014}
\bibinfo{author}{T. Wehling}, \bibinfo{author}{A. Black-Schaffer}, \ and\
  \bibinfo{author}{A. Balatsky},
\newblock \bibinfo{title}{{Dirac materials}}.
\newblock \emph{\bibinfo{journal}{Adv. Phys.}}
  \textbf{\bibinfo{volume}{63}}, \bibinfo{pages}{1} (\bibinfo{year}{2014}).

\bibitem{Yeats2015}
\bibinfo{author}{A.~L. Yeats}, \bibinfo{author}{Y. Pan}, \bibinfo{author}{A. Richardella}, \bibinfo{author}{P.~J. Mintun}, \bibinfo{author}{N. Samarth}, \ and\ \bibinfo{author}{D.~D. Awschalom},
\newblock \bibinfo{title}{{Persistent optical gating of a topological
  insulator}}.
\newblock \emph{\bibinfo{journal}{Science Adv.}}
  \textbf{\bibinfo{volume}{1}}, \bibinfo{pages}{e1500640}
  (\bibinfo{year}{2015}).

\bibitem{Hou2020}
\bibinfo{author}{Y. Hou}, \bibinfo{author}{R. Xiao}, \bibinfo{author}{S. Li}, \bibinfo{author}{L. Wang}, \ and\ \bibinfo{author}{D. Yu},
\newblock \bibinfo{title}{{Nonlocal Chemical Potential Modulation in Topological Insulators Enabled by Highly Mobile Trapped Charges}}.
\newblock \emph{\bibinfo{journal}{ACS Appl. Electron. Mater.}} \textbf{\bibinfo{volume}{2}}, \bibinfo{pages}{3436} (\bibinfo{year}{2020}).

\bibitem{Xie2021}
\bibinfo{author}{Xie, F.}, \bibinfo{author}{L. Zhen}, \bibinfo{author}{S. Zhang}, \bibinfo{author}{T. Wang}, \bibinfo{author}{S. Miao}, \bibinfo{author}{Z. Song}, \bibinfo{author}{Y. Ying}, \bibinfo{author}{X.~C. Pan}, \bibinfo{author}{M. Long}, \bibinfo{author}{M. Zhang}, \emph{et~al.},
\newblock \bibinfo{title}{{Reversible engineering of topological insulator
  surface state conductivity through optical excitation}}.
\newblock \emph{\bibinfo{journal}{Nanotechnol.}}
  \textbf{\bibinfo{volume}{32}}, \bibinfo{pages}{17LT01}
  (\bibinfo{year}{2021}).
  
\bibitem{Zheng2015}
\bibinfo{author}{K. Zheng}, \bibinfo{author}{L.~B. Luo}, \bibinfo{author}{T.~F. Zhang}, \bibinfo{author}{Y.~H. Liu}, \bibinfo{author}{Y.~Q. Yu}, \bibinfo{author}{R. Lu}, \bibinfo{author}{H.~L. Qiu}, \bibinfo{author}{Z.~J. Li}, \bibinfo{author}{K. Zheng}, \ and\ \bibinfo{author}{J.~C. Andrew Huang},
\newblock \bibinfo{title}{{Optoelectronic characteristics of a near infrared light photodetector
	based on a topological insulator Sb$_2$Te$_3$ film}}.
\newblock \emph{\bibinfo{journal}{J. Mater. Chem. C.}}
  \textbf{\bibinfo{volume}{3}}, \bibinfo{pages}{9154}
  (\bibinfo{year}{2015}).

\bibitem{Pirralho2019}
\bibinfo{author}{M.~J.~P. Pirralho}, \bibinfo{author}{M.~L. Peres}, \bibinfo{author}{C.~I. Fornari}, \bibinfo {author} {D.~P.~A. Holgado}, \bibinfo {author} {F.~S. Pena},
  \bibinfo{author} {S. Nakamatsu}, \bibinfo {author} {P.~H.~O. Rappl}, \bibinfo {author}{E. Abramof}, \ and\ \bibinfo {author}
  {D.~A.~W. Soares},
\newblock \bibinfo{title}{{Investigation of photoconductive effect on
  ${\mathrm{Bi}}_{2}{\mathrm{Te}}_{3}$ epitaxial film}}.
\newblock \emph{\bibinfo{journal}{Appl. Phys. Lett.}}
  \textbf{\bibinfo{volume}{114}}, \bibinfo{pages}{112101}
  (\bibinfo{year}{2019}).

\bibitem{Kang2019}
\bibinfo{author}{T.~T. Kang}, \ and\ \bibinfo{author}{P.~P. Chen},
\newblock \bibinfo{title}{{Bi$_2$Te$_3$ photoconductive detector under weak light}}.
\newblock \emph{\bibinfo{journal}{J. Appl. Phys.}}
  \textbf{\bibinfo{volume}{126}}, \bibinfo{pages}{083103}
  (\bibinfo{year}{2019}).

\bibitem{Tian2017}
\bibinfo{author}{J. Tian}, \bibinfo{author}{S. Hong},
  \bibinfo{author}{I. Miotkowski}, \bibinfo{author}{S. Datta}, \ and\
  \bibinfo{author}{Y.~P. Chen},
\newblock \bibinfo{title}{{Observation of current-induced, long-lived
  persistent spin polarization in a topological insulator: A rechargeable spin
  battery}}.
\newblock \emph{\bibinfo{journal}{Science Adv.}}
  \textbf{\bibinfo{volume}{3}}, \bibinfo{pages}{e1602531}
  (\bibinfo{year}{2017}).

\bibitem{Li2017}
\bibinfo {author} {M. Li}, \bibinfo {author} {Z. Wang},
  \bibinfo {author} {L. Yang}, \bibinfo
  {author} {D. Li}, \bibinfo {author}
  {Q.~R. Yao}, \bibinfo {author} {G.~H. Rao}, \bibinfo {author} {X.P.~A. Gao}, \ and\ \bibinfo {author} {Z. Zhang},
\newblock \bibinfo{title}{{Electron delocalization and relaxation behavior in
  Cu-doped $\mathrm{B}{\mathrm{i}}_{2}\mathrm{S}{\mathrm{e}}_{3}$ films}}.
\newblock \emph{\bibinfo{journal}{Phys. Rev. B}} \textbf{\bibinfo{volume}{96}},
  \bibinfo{pages}{075152} (\bibinfo{year}{2017}).

\bibitem{Uchiyama2023}
\bibinfo{author}{T. Uchiyama}, \bibinfo{author}{H. Goto}, \bibinfo{author}{E. Uesugi}, \bibinfo{author}{A. Takai}, \bibinfo{author}{L. Zhi}, \bibinfo{author}{A. Miura}, \bibinfo{author}{S. Hamao}, \bibinfo{author}{R. Eguchi}, \bibinfo{author}{H. Ota}, \bibinfo{author}{K. Sugimoto}, \emph{et~al.}
\newblock \bibinfo{title}{{Semiconductor–metal transition in Bi$_2$Se$_3$ caused by impurity doping}}.
\newblock \emph{\bibinfo{journal}{Sci. Rep.}} \textbf{\bibinfo{volume}{13}},
  \bibinfo{pages}{537} (\bibinfo{year}{2023}).

\bibitem{Ishida2015}
\bibinfo{author}{Y. Ishida}, \bibinfo{author}{T. Otsu}, \bibinfo{author}{T. Shimada}, \bibinfo{author}{M. Okawa}, \bibinfo{author}{Y. Kobayashi}, \bibinfo{author}{F. Iga}, \bibinfo{author}{T. Takabatake}, \ and\ \bibinfo{author}{S. Shin},
\newblock \bibinfo{title}{{Emergent photovoltage on SmB$_6$ surface upon
  bulk-gap evolution revealed by pump-and-probe photoemission spectroscopy}}.
\newblock \emph{\bibinfo{journal}{Sci. Rep.}}
  \textbf{\bibinfo{volume}{5}}, \bibinfo{pages}{8160} (\bibinfo{year}{2015}).
  
\bibitem{Neupane2015}
\bibinfo{author}{M. Neupane}, \bibinfo{author}{S.~Y. Xu}, \bibinfo{author}{Y. Ishida}, \bibinfo{author}{S. Jia}, \bibinfo{author}{B.~M. Fregoso}, \bibinfo{author}{C. Liu}, \bibinfo{author}{I. Belopolski}, \bibinfo{author}{G. Bian}, \bibinfo{author}{N. Alidoust}, \bibinfo{author}{T. Durakiewicz}, \emph{et~al.},
\newblock \bibinfo{title}{{Gigantic Surface Lifetime of an Intrinsic
  Topological Insulator}}.
\newblock \emph{\bibinfo{journal}{Phys. Rev. Lett.}}
  \textbf{\bibinfo{volume}{115}}, \bibinfo{pages}{116801}
  (\bibinfo{year}{2015}).

\bibitem{Sumida2017}
\bibinfo{author}{K. Sumida}, \bibinfo{author}{Y. Ishida}, \bibinfo{author}{S. Zhu}, \bibinfo{author}{M. Ye}, \bibinfo{author}{A. Pertsova}, \bibinfo{author}{C. Triola}, \bibinfo{author}{K.~A. Kokh}, \bibinfo{author}{O.~E. Tereshchenko}, \bibinfo{author}{A.~V. Balatsky}, \bibinfo{author}{S. Shin}, \ and\ \bibinfo{author}{A. Kimura},
\newblock \bibinfo{title}{{Prolonged duration of nonequilibrated Dirac fermions
  in neutral topological insulators}}.
\newblock \emph{\bibinfo{journal}{Sci. Rep.}}
  \textbf{\bibinfo{volume}{7}}, \bibinfo{pages}{14080}
  (\bibinfo{year}{2017}).

\bibitem{Yoshikawa2018}
\bibinfo {author} {T. Yoshikawa}, \bibinfo {author} {Y. Ishida}, \bibinfo {author} {K. Sumida},
  \bibinfo {author} {J. Chen}, \bibinfo
  {author} {K.~A. Kokh}, \bibinfo
  {author} {O.~E. Tereshchenko}, \bibinfo
  {author} {S. Shin}, \ and\ \bibinfo
  {author} {A. Kimura},
\newblock \bibinfo{title}{{Enhanced photovoltage on the surface of topological
  insulator via optical aging}}.
\newblock \emph{\bibinfo{journal}{Appl. Phys. Lett.}}
  \textbf{\bibinfo{volume}{112}}, \bibinfo{pages}{192104}
  (\bibinfo{year}{2018}).

\bibitem{Papalazarou2018}
\bibinfo{author}{E. Papalazarou}, \bibinfo{author}{L. Khalil}, \bibinfo{author}{M. Caputo}, \bibinfo{author}{L. Perfetti}, \bibinfo{author}{N. Nilforoushan}, \bibinfo{author}{H. Deng}, \bibinfo{author}{Z. Chen}, \bibinfo{author}{S. Zhao}, \bibinfo{author}{A. Taleb-Ibrahimi}, \bibinfo{author}{M. Konczykowski}, \emph{et~al.},
\newblock \bibinfo{title}{{Unraveling the Dirac fermion dynamics of the
  bulk-insulating topological system
  ${\mathrm{Bi}}_{2}{\mathrm{Te}}_{2}\mathrm{Se}$}}.
\newblock \emph{\bibinfo{journal}{Phys. Rev. Mater.}}
  \textbf{\bibinfo{volume}{2}}, \bibinfo{pages}{104202}
  (\bibinfo{year}{2018}).

\bibitem{Yoshikawa2019}
\bibinfo{author}{T. Yoshikawa}, \bibinfo{author}{K. Sumida}, \bibinfo{author}{Y. Ishida}, \bibinfo{author}{J. Chen}, \bibinfo{author}{M. Nurmamat}, \bibinfo{author}{K. Akiba}, \bibinfo{author}{A. Miyake}, \bibinfo{author}{M. Tokunaga}, \bibinfo{author}{K.~A. Kokh}, \bibinfo{author}{O.~E. Tereshchenko}, \emph{et~al.},
\newblock \bibinfo{title}{{Bidirectional surface photovoltage on a topological
  insulator}}.
\newblock \emph{\bibinfo{journal}{Phys. Rev. B}}
  \textbf{\bibinfo{volume}{100}}, \bibinfo{pages}{165311}
  (\bibinfo{year}{2019}).

\bibitem{Ciocys2020}
\bibinfo{author}{S. Ciocys}, \bibinfo{author}{T. Morimoto}, \bibinfo{author}{R. Mori}, \bibinfo{author}{K. Gotlieb}, \bibinfo{author}{Z. Hussain}, \bibinfo{author}{J.~G. Analytis}, \bibinfo{author}{J.~E. Moore}, \ and\ \bibinfo{author}{A. Lanzara},
\newblock \bibinfo{title}{{Manipulating long-lived topological surface
  photovoltage in bulk-insulating topological insulators
  ${\mathrm{Bi}}_{2}{\mathrm{Se}}_{3}$ and
  ${\mathrm{Bi}}_{2}{\mathrm{Te}}_{3}$}}.
\newblock \emph{\bibinfo{journal}{npj Quantum Mater.}}
  \textbf{\bibinfo{volume}{5}}, \bibinfo{pages}{16} (\bibinfo{year}{2020}).
  
\bibitem{Eremeev2010}
\bibinfo{author}{S.~V. Eremeev}, \bibinfo{author}{Y.~M. Koroteev}, \ and\
  \bibinfo{author}{E.~V. Chulkov},
\newblock \bibinfo{title}{{Ternary thallium-based semimetal chalcogenides
  Tl-V-VI$_2$ as a new class of three-dimensional topological insulators}}.
\newblock \emph{\bibinfo{journal}{JETP Lett.}} \textbf{\bibinfo{volume}{91}},
  \bibinfo{pages}{594} (\bibinfo{year}{2010}).

\bibitem{Lin2010}
\bibinfo{author}{H. Lin}, \bibinfo{author}{R.~S. Markiewicz}, \bibinfo{author}{L.~A. Wray}, \bibinfo{author}{L. Fu}, \bibinfo{author}{M.~Z. Hasan}, \ and\ \bibinfo{author}{A. Bansil},
\newblock \bibinfo{title}{{Single-Dirac-Cone Topological Surface States in the
  ${\mathrm{TlBiSe}}_{2}$ Class of Topological Semiconductors}}.
\newblock \emph{\bibinfo{journal}{Phys. Rev. Lett.}}
  \textbf{\bibinfo{volume}{105}}, \bibinfo{pages}{036404}
  (\bibinfo{year}{2010}).

\bibitem{Yan2010}
\bibinfo{author}{B. Yan}, \bibinfo{author}{C.-X. Liu}, \bibinfo{author}{H.-J. Zhang}, \bibinfo{author}{C.-Y. Yam}, \bibinfo{author}{X.-L. Qi}, \bibinfo{author}{T. Frauenheim}, \ and\ \bibinfo{author}{S.-C. Zhang},
\newblock \bibinfo{title}{{Theoretical prediction of topological insulators in
  thallium-based {III}-V-{VI}$_2$ ternary chalcogenides}}.
\newblock \emph{\bibinfo{journal}{{EPL} (Europhys. Lett.)}}
  \textbf{\bibinfo{volume}{90}}, \bibinfo{pages}{37002}
  (\bibinfo{year}{2010}).

\bibitem{Sato2010}
\bibinfo{author}{T. Sato}, \bibinfo{author}{K. Segawa}, \bibinfo{author}{H. Guo}, \bibinfo{author}{K. Sugawara}, \bibinfo{author}{S. Souma}, \bibinfo{author}{T. Takahashi}, \ and\ \bibinfo{author}{Y. Ando},
\newblock \bibinfo{title}{{Direct Evidence for the Dirac-Cone Topological
  Surface States in the Ternary Chalcogenide ${\mathrm{TlBiSe}}_{2}$}}.
\newblock \emph{\bibinfo{journal}{Phys. Rev. Lett.}}
  \textbf{\bibinfo{volume}{105}}, \bibinfo{pages}{136802}
  (\bibinfo{year}{2010}).

\bibitem{Kuroda2010}
\bibinfo{author}{K. Kuroda}, \bibinfo{author}{M. Ye}, \bibinfo{author}{A. Kimura}, \bibinfo{author}{S.~V. Eremeev}, \bibinfo{author}{E.~E. Krasovskii}, \bibinfo{author}{E.~V. Chulkov}, \bibinfo{author}{Y. Ueda}, \bibinfo{author}{K. Miyamoto}, \bibinfo{author}{T. Okuda}, \bibinfo{author}{K. Shimada}, \emph{et~al.},
\newblock \bibinfo{title}{{Experimental Realization of a Three-Dimensional
  Topological Insulator Phase in Ternary Chalcogenide
  ${\mathrm{TlBiSe}}_{2}$}}.
\newblock \emph{\bibinfo{journal}{Phys. Rev. Lett.}}
  \textbf{\bibinfo{volume}{105}}, \bibinfo{pages}{146801}
  (\bibinfo{year}{2010}).

\bibitem{Kuroda2015}
\bibinfo {author} {K. Kuroda}, \bibinfo {author} {G. Eguchi},  \bibinfo {author} {K. Shirai}, \bibinfo
  {author} {M. Shiraishi}, \bibinfo {author} {M. Ye}, \bibinfo {author}
  {K. Miyamoto}, \bibinfo {author}{T. Okuda}, \bibinfo {author}
  {S. Ueda}, \bibinfo {author}{M. Arita}, \bibinfo {author}
  {H. Namatame}, \emph{et~al.},
\newblock \bibinfo{title}{{Tunable spin current due to bulk insulating property
  in the topological insulator
  ${\mathrm{Tl}}_{1\ensuremath{-}x}{\mathrm{Bi}}_{1+x}{\mathrm{Se}}_{2\ensuremath{-}\ensuremath{\delta}}$}}.
\newblock \emph{\bibinfo{journal}{Phys. Rev. B}} \textbf{\bibinfo{volume}{91}},
  \bibinfo{pages}{205306} (\bibinfo{year}{2015}).

\bibitem{Ishida2014}
\bibinfo{author}{Y. Ishida}, \bibinfo{author}{T. Togashi}, \bibinfo{author}{K. Yamamoto}, \bibinfo{author}{M. Tanaka}, \bibinfo{author}{T. Kiss}, \bibinfo{author}{T. Otsu}, \bibinfo{author}{Y. Kobayashi}, \ and\ \bibinfo{author}{S. Shin},
\newblock \bibinfo{title}{{Time-resolved photoemission apparatus achieving
  sub-20-meV energy resolution and high stability}}.
\newblock \emph{\bibinfo{journal}{Rev. of Sci. Instrum.}}
  \textbf{\bibinfo{volume}{85}}, \bibinfo{pages}{123904}
  (\bibinfo{year}{2014}).

\bibitem{Eguchi2014}
\bibinfo{author}{G. Eguchi}, \bibinfo{author}{K. Kuroda},
  \bibinfo{author}{K. Shirai}, \bibinfo{author}{A. Kimura}, \ and\
  \bibinfo{author}{M. Shiraishi},
\newblock \bibinfo{title}{{Surface Shubnikov-de Haas oscillations and nonzero
  Berry phases of the topological hole conduction in
  ${\mathrm{Tl}}_{1\ensuremath{-}x}{\mathrm{Bi}}_{1+x}{\mathrm{Se}}_{2}$}}.
\newblock \emph{\bibinfo{journal}{Phys. Rev. B}} \textbf{\bibinfo{volume}{90}},
  \bibinfo{pages}{201307} (\bibinfo{year}{2014}).

\bibitem{Eguchi2015}
\bibinfo{author}{G. Eguchi}, \bibinfo{author}{K. Kuroda}, \bibinfo{author}{K. Shirai}, \bibinfo{author}{Y. Ando}, \bibinfo{author}{T. Shinjo}, \bibinfo{author}{A. Kimura}, \ and\ \bibinfo{author}{M. Shiraishi},
\newblock \bibinfo{title}{{Precise determination of two-carrier transport
  properties in the topological insulator ${\mathrm{TlBiSe}}_{2}$}}.
\newblock \emph{\bibinfo{journal}{Phys. Rev. B}} \textbf{\bibinfo{volume}{91}},
  \bibinfo{pages}{235117} (\bibinfo{year}{2015}).

\bibitem{Eguchi2019}
\bibinfo{author}{G. Eguchi}, \ and\ \bibinfo{author}{S. Paschen},
\newblock \bibinfo{title}{{Robust scheme for magnetotransport analysis in
  topological insulators}}.
\newblock \emph{\bibinfo{journal}{Phys. Rev. B}} \textbf{\bibinfo{volume}{99}},
  \bibinfo{pages}{165128} (\bibinfo{year}{2019}).

\bibitem{Hsieh2009}
\bibinfo{author}{D. Hsieh}, \bibinfo{author}{Y. Xia}, \bibinfo{author}{D. Qian}, \bibinfo{author}{L. Wray}, \bibinfo{author}{J.~H. Dil}, \bibinfo{author}{F. Meier}, \bibinfo{author}{J. Osterwalder}, \bibinfo{author}{L. Patthey}, \bibinfo{author}{J.~G. Checkelsky}, \bibinfo{author}{N.~P. Ong}, \emph{et~al.},
\newblock \bibinfo{title}{{A tunable topological insulator in the spin helical
  Dirac transport regime}}.
\newblock \emph{\bibinfo{journal}{Nature}} \textbf{\bibinfo{volume}{460}},
  \bibinfo{pages}{1101} (\bibinfo{year}{2009}).

\bibitem{Checkelsky2011}
\bibinfo{author}{J.~G. Checkelsky}, \bibinfo{author}{Y.~S. Hor},
  \bibinfo{author}{R.~J. Cava}, \ and\ \bibinfo{author}{N.~P. Ong},
\newblock \bibinfo{title}{{Bulk Band Gap and Surface State Conduction Observed
  in Voltage-Tuned Crystals of the Topological Insulator
  ${\mathrm{Bi}}_{2}{\mathrm{Se}}_{3}$}}.
\newblock \emph{\bibinfo{journal}{Phys. Rev. Lett.}}
  \textbf{\bibinfo{volume}{106}}, \bibinfo{pages}{196801}
  (\bibinfo{year}{2011}).

\bibitem{Frantzeskakis2017}
\bibinfo {author} {E. Frantzeskakis}, \bibinfo {author} {S.~V. Ramankutty}, \bibinfo {author} {N. de~Jong},
  \bibinfo {author} {Y.~K. Huang}, \bibinfo{author} {Y. Pan}, \bibinfo {author}{A. Tytarenko}, \bibinfo {author}{M. Radovic}, \bibinfo {author} {N.~C. Plumb}, \bibinfo {author} {M. Shi}, \bibinfo {author} {A. Varykhalov}, \emph{et~al.},
\newblock \bibinfo{title}{{Trigger of the Ubiquitous Surface Band Bending in 3D
  Topological Insulators}}.
\newblock \emph{\bibinfo{journal}{Phys. Rev. X}} \textbf{\bibinfo{volume}{7}},
  \bibinfo{pages}{041041} (\bibinfo{year}{2017}).

\bibitem{Taskin2011}
\bibinfo{author}{A.~A. Taskin}, \bibinfo{author}{Z. Ren},
  \bibinfo{author}{S. Sasaki}, \bibinfo{author}{K. Segawa}, \ and\
  \bibinfo{author}{Y. Ando},
\newblock \bibinfo{title}{{Observation of Dirac Holes and Electrons in a
  Topological Insulator}}.
\newblock \emph{\bibinfo{journal}{Phys. Rev. Lett.}}
  \textbf{\bibinfo{volume}{107}}, \bibinfo{pages}{016801}
  (\bibinfo{year}{2011}).

\bibitem{Phillips1996}
\bibinfo{author}{J.~C. Phillips},
\newblock \bibinfo{title}{{Stretched exponential relaxation in molecular and
  electronic glasses}}.
\newblock \emph{\bibinfo{journal}{Rep. on Prog. in Phys.}}
  \textbf{\bibinfo{volume}{59}}, \bibinfo{pages}{1133}
  (\bibinfo{year}{1996}).

\bibitem{Sze2007}
\bibinfo{author}{S.~M. Sze}, \ and\ \bibinfo{author}{K.~N. Kwok},
\newblock \emph{\bibinfo{title}{{Physics of semiconductor devices}}}
  (\bibinfo{publisher}{Hoboken, N.J: Wiley-Interscience},
  \bibinfo{year}{2007}).

\bibitem{Fu2013}
\bibinfo {author} {Y.-S. Fu}, \bibinfo {author} {T. Hanaguri}, \bibinfo {author} {S. Yamamoto}, \bibinfo {author} {K. Igarashi}, \bibinfo {author} {H. Takagi}, \ and\ \bibinfo {author} {T. Sasagawa},
\newblock \bibinfo{title}{{Memory Effect in a Topological Surface State of
  ${\mathrm{Bi}}_{2}{\mathrm{Te}}_{2}\mathrm{Se}$}}.
\newblock \emph{\bibinfo{journal}{ACS Nano}} \textbf{\bibinfo{volume}{7}},
  \bibinfo{pages}{4105} (\bibinfo{year}{2013}).

\bibitem{Kalkan2015}
\bibinfo{author}{N. Kalkan}, \ and\ \bibinfo{author}{H. Bas},
\newblock \bibinfo{title}{{Electrical and Switching Properties of TlBiSe$_2$
  Chalcogenide Compounds}}.
\newblock \emph{\bibinfo{journal}{J. of Electron. Mater.}}
  \textbf{\bibinfo{volume}{44}}, \bibinfo{pages}{4387}
  (\bibinfo{year}{2015}).

\bibitem{Lang1977}
\bibinfo{author}{D.~V. Lang}, \ and\ \bibinfo{author}{R.~A. Logan},
\newblock \bibinfo{title}{{Large-Lattice-Relaxation Model for Persistent
  Photoconductivity in Compound Semiconductors}}.
\newblock \emph{\bibinfo{journal}{Phys. Rev. Lett.}}
  \textbf{\bibinfo{volume}{39}}, \bibinfo{pages}{635} (\bibinfo{year}{1977}).

\bibitem{Zukotynski1987}
\bibinfo{author}{S. Zukotynski}, \bibinfo{author}{P.~C.~H. Ng}, \ and\ \bibinfo{author}{A.~J. Pindor},
\newblock \bibinfo{title}{{Persistent photoconductivity in Si-doped Al$_x$Ga$_{1-x}$As}}.
\newblock \emph{\bibinfo{journal}{Phys. Rev. Lett.}}
  \textbf{\bibinfo{volume}{59}}, \bibinfo{pages}{2810} (\bibinfo{year}{1987}).
   
\bibitem{Scalvi2020} 
\bibinfo{author}{L.~V.~A. Scalvi}, \ and\ \bibinfo{author}{C.~F. Bueno},
\newblock \bibinfo{title}{{Transient decay of photoinduced current in semiconductors and heterostructures}}.
\newblock \emph{\bibinfo{journal}{J. Phys. D: Appl. Phys.}}
  \textbf{\bibinfo{volume}{53}}, \bibinfo{pages}{033001} (\bibinfo{year}{2020}).

\bibitem{Pielmeier2015}
\bibinfo{author}{F. Pielmeier}, \bibinfo{author}{G. Landolt}, \bibinfo{author}{B. Slomski}, \bibinfo{author}{S. Muff}, \bibinfo{author}{J. Berwanger}, \bibinfo{author}{A. Eich}, \bibinfo{author}{A.~A. Khajetoorians}, \bibinfo{author}{J. Wiebe}, \bibinfo{author}{Z.~S. Aliev}, \ and\ \bibinfo{author}{M.~B. Babanly},
\newblock \bibinfo{title}{{Response of the topological surface state to surface disorder in TlBiSe$_2$}}.
\newblock \emph{\bibinfo{journal}{New J. of Phys.}} \textbf{\bibinfo{volume}{17}},
  \bibinfo{pages}{023067} (\bibinfo{year}{2015}). 
 
\bibitem{Singh2016} 
\bibinfo{author}{B. Singh}, \bibinfo{author}{H. Lin}, \bibinfo{author}{R. Prasad}, \ and\ \bibinfo{author}{A. Bansil},
\newblock \bibinfo{title}{{Role of surface termination in realizing well-isolated topological surface states within the bulk band gap in ${\mathrm{TlBiSe}}_{2}$ and ${\mathrm{TlBiTe}}_{2}$}}.
\newblock \emph{\bibinfo{journal}{Phys. Rev. B}}
\textbf{\bibinfo{volume}{93}}, \bibinfo{pages}{085113} (\bibinfo{year}{2016}).

\bibitem{Konorov2006}
\bibinfo{author}{P.~P. Konorov}, \bibinfo{author}{A. Yafyasov}, \ and\
  \bibinfo{author}{V. Bogevolnov},
\newblock \emph{\bibinfo{title}{Field Effect in Semiconductor-electrolyte
  Interfaces: Application to Investigations of Electronic Properties of
  Semiconductor Surfaces}} (\bibinfo{publisher}{Princeton University Press},
  \bibinfo{year}{2006}).

\bibitem{Mitsas1992}
\bibinfo{author}{C. Mitsas}, \ and\ \bibinfo{author}{D. Siapkas},
\newblock \bibinfo{title}{{Phonon and electronic properties of TlBiSe$_2$ thin
  films}}.
\newblock \emph{\bibinfo{journal}{Solid State Commun.}}
  \textbf{\bibinfo{volume}{83}}, \bibinfo{pages}{857} (\bibinfo{year}{1992}).

\bibitem{Mitsas1993}
\bibinfo{author}{C.~L. Mitsas}, \bibinfo{author}{E.~K. Polychroniadis}, \ and\
  \bibinfo{author}{D.~I. Siapkas},
\newblock \bibinfo{title}{{Structural dependence of the optical absorption in
  {TlBiSe}$_2$ thin films near the fundamental absorption edge}}.
\newblock \emph{\bibinfo{journal}{Semicond. Science and Technol.}}
  \textbf{\bibinfo{volume}{8}}, \bibinfo{pages}{S356} (\bibinfo{year}{1993}).

\bibitem{Queisser1985}
\bibinfo{author}{H.~J. Queisser},
\newblock \bibinfo{title}{{Nonexponential Relaxation of Conductance near
  Semiconductor Interfaces}}.
\newblock \emph{\bibinfo{journal}{Phys. Rev. Lett.}}
  \textbf{\bibinfo{volume}{54}}, \bibinfo{pages}{234}
  (\bibinfo{year}{1985}).

\bibitem{Queisser1986}
\bibinfo{author}{H.~J. Queisser}, \ and\ \bibinfo{author}{D.~E. Theodorou},
\newblock \bibinfo{title}{{Decay kinetics of persistent photoconductivity in
  semiconductors}}.
\newblock \emph{\bibinfo{journal}{Phys. Rev. B}} \textbf{\bibinfo{volume}{33}},
  \bibinfo{pages}{4027} (\bibinfo{year}{1986}).

\bibitem{Friedensen2017}
\bibinfo{author}{S. Friedensen}, \bibinfo{author}{J.~T. Mlack}, \ and\
  \bibinfo{author}{M. Drndic},
\newblock \bibinfo{title}{{Materials analysis and focused ion beam
  nanofabrication of topological insulator
  ${\mathrm{Bi}}_{2}{\mathrm{Se}}_{3}$}}.
\newblock \emph{\bibinfo{journal}{Sci. Rep.}}
  \textbf{\bibinfo{volume}{7}}, \bibinfo{pages}{13466}
  (\bibinfo{year}{2017}).
 
\bibitem{Ren2011}
\bibinfo{author}{Z. Ren}, \bibinfo{author}{A.~A. Taskin}, \bibinfo{author}{S. Sasaki}, \bibinfo{author}{K. Segawa}, \ and\ \bibinfo{author}{Y. Ando},
\newblock \bibinfo{title}{{Observations of two-dimensional quantum oscillations and ambipolar transport in the topological insulator Bi$_2$Se$_3$ achieved by Cd doping}}.
\newblock \emph{\bibinfo{journal}{Phys. Rev. B}} \textbf{\bibinfo{volume}{84}},
  \bibinfo{pages}{075316} (\bibinfo{year}{2011}).

\bibitem{Kushwaha2016}
\bibinfo{author}{S.~K. Kushwaha}, \bibinfo{author}{I. Pletikosic}, \bibinfo{author}{T. Liang}, \bibinfo{author}{A. Gyenis}, \bibinfo{author}{S.~H. Lapidus}, \bibinfo{author}{Y. Tian}, \bibinfo{author}{H. Zhao}, \bibinfo{author}{K.~S. Burch}, \bibinfo{author}{J. Lin}, \bibinfo{author}{W. Wang}, \emph{et~al.},
\newblock \bibinfo{title}{{Sn-doped
  ${\mathrm{Bi}}_{1.1}{\mathrm{Sb}}_{0.9}{\mathrm{Te}}_{2}{\mathrm{S}}$ bulk
  crystal topological insulator with excellent properties}}.
\newblock \emph{\bibinfo{journal}{Nat. Commun.}}
  \textbf{\bibinfo{volume}{7}}, \bibinfo{pages}{11456}
  (\bibinfo{year}{2016}).

\bibitem{Borgwardt2016}
\bibinfo {author} {N. Borgwardt}, \bibinfo {author} {J. Lux},
  \bibinfo {author} {I. Vergara}, \bibinfo
  {author} {Z. Wang}, \bibinfo {author}
  {A.~A. Taskin}, \bibinfo {author}
  {K. Segawa}, \bibinfo {author}
  {P.~H.~M. van Loosdrecht}, \bibinfo {author}
  {Y. Ando}, \bibinfo {author}
  {A. Rosch}, \ and\ \bibinfo {author}
  {M. Gr\"uninger},
\newblock \bibinfo{title}{{Self-organized charge puddles in a three-dimensional
  topological material}}.
\newblock \emph{\bibinfo{journal}{Phys. Rev. B}} \textbf{\bibinfo{volume}{93}},
  \bibinfo{pages}{245149} (\bibinfo{year}{2016}).

\bibitem{Le2021}
\bibinfo{author}{T. Le}, \bibinfo{author}{Q. Ye}, \bibinfo{author}{C. Chen}, \bibinfo{author}{L. Yin}, \bibinfo{author}{D. Zhang}, \bibinfo{author}{X. Wang}, \ and\ \bibinfo{author}{X. Lu},
\newblock \bibinfo{title}{{Erasable superconductivity in topological insulator Bi$_2$Se$_3$ induced by voltage pulse}}.
\newblock \emph{\bibinfo{journal}{Adv. Quantum Technol.}} \textbf{\bibinfo{volume}{4}},
  \bibinfo{pages}{2100067} (\bibinfo{year}{2021}).

\bibitem{Pirie2021}
\bibinfo{author}{H. Pirie}, \bibinfo{author}{Y. Liu}, \bibinfo{author}{A. Soumyanarayanan}, \bibinfo{author}{P. Chen}, \bibinfo{author}{Y. He}, \bibinfo{author}{M.~M. Yee}, \bibinfo{author}{P.~F.~S. Rosa}, \bibinfo{author}{J.~D. Thompson}, \bibinfo{author}{D.-J. Kim}, \bibinfo{author}{Z. Fisk}, \emph{et~al.},
\newblock \bibinfo{title}{{Imaging emergent heavy Dirac fermions of a topological Kondo insulator}}.
\newblock \emph{\bibinfo{journal}{Nat. Phys.}} \textbf{\bibinfo{volume}{16}},
  \bibinfo{pages}{52} (\bibinfo{year}{2020}).

\end{thebibliography}


\newpage


\clearpage


\clearpage


\clearpage


\clearpage


\end{document}